\documentclass[aps,11pt,superscriptaddress,nofootinbib,notitlepage,showkeys]{revtex4-2}

\pdfoutput=1
\usepackage[T1]{fontenc}
\usepackage{comment}
\usepackage{graphicx}
\usepackage{bm}
\usepackage{amsfonts,mathtools,amssymb,stmaryrd}
\usepackage{color}
\usepackage[breaklinks]{hyperref}
\usepackage{epstopdf, epsfig}
\usepackage{natbib}
\usepackage{grffile}
\usepackage{amsthm}
\usepackage{amssymb}
\usepackage{amsmath}
\usepackage{bbm}
\usepackage{MnSymbol}
\newcommand{\mA}{\mathcal{A}}

\newcommand{\mN}{{\mathcal N}}

\newcommand{\mS}{\mathcal{S}}

\newcommand{\bx}{{\bf x}}
\newcommand{\bu}{{\bf u}}

\newcommand{\mX}{{\mathcal X}}

\newcommand{\be}{\begin{equation}}
\newcommand{\ee}{\end{equation}}
\newcommand{\bsp}{\begin{split}}
\newcommand{\esp}{\end{split}}

\newcommand{\bi}{\begin{itemize}}
\newcommand{\ei}{\end{itemize}}

\newcommand{\av}[1]{\left\langle#1\right\rangle}
\newcommand{\Pe}{\mathrm{Pe}}

\newcommand{\corr}{\color{black}}
\newcommand{\rroc}{\color{black}}

\newcommand{\ito}{It\^o}

\begin{document}
\title{From zero-mode intermittency to hidden symmetry in random scalar advection}

\author{Simon Thalabard}
\affiliation{Institut de Physique de Nice, Universit\'e C\^ote d'Azur CNRS - UMR 7010, 17 rue Julien Laupr\^etre, 06200 Nice, France}
\email{simon.thalabard@univ-cotedazur.fr}

\author{Alexei A. Mailybaev} 
\affiliation{Instituto de Matem\'atica Pura e Aplicada -- IMPA, Rio de Janeiro, Brazil}
\email{alexei@impa.br}

\date{\today}
\begin{abstract}
The statistical behavior of scalars passively advected by random flows exhibits intermittency in the form of anomalous multiscaling, in many ways similar to the patterns commonly observed in incompressible high-Reynolds fluids. This similarity suggests a generic dynamical mechanism underlying intermittency, though its specific nature remains unclear. Scalar turbulence is framed in a linear setting that points towards a zero-mode scenario connecting anomalous scaling to the presence of statistical conservation laws; the duality is fully substantiated within Kraichnan theory of random flows. However, extending the zero-mode scenario to nonlinear settings faces formidable technical challenges.
Here, we revisit the scalar problem in the light of a hidden symmetry scenario introduced in recent deterministic turbulence studies addressing the Sabra shell model and the Navier-Stokes equations. Hidden symmetry uses a rescaling strategy based entirely on symmetry considerations, transforming the original dynamics into a rescaled (hidden) system; \corr It yields the universality of Kolmogorov multipliers and ultimately identifies the scaling exponents as the eigenvalues of Perron-Frobenius operators. \rroc
Considering a minimal shell model of scalar advection of the Kraichnan type that was previously studied by Biferale \& Wirth, the present work extends the hidden symmetry approach to a stochastic setting, in order to explicitly contrast it with the zero-mode scenario. 
\corr Our study indicates that the zero-mode and the multiplicative scenarios are intrinsically related. While the zero-mode approach solves the eigenvalue problem for $p^{th}$ order correlation functions, Perron-Frobenius (multiplicative) scenario defines a similar eigenvalue problem in terms of $p^{th}$ order measures. 
For systems of the Kraichnan type,  the first approach provides a  quantitative chararacterization of intermittency,
while the second  approach highlights the universal connection between the scalar case and a larger class of hydrodynamic models. \rroc
\end{abstract}

\maketitle
%\tableofcontents
%\newpage
%##############################################
%INTRODUCTION-DEFINITION
\section{Introduction}
\label{sec:intro}
 A hallmark feature of turbulent environments is their apparent lack of statistical self-similarity,  a feature commonly referred to as spatial intermittency. While such phenomenology is successfully captured by  statistical frameworks involving multifractal random processes of 
various complexities 
\cite{frisch1980fully,frisch1985singularity,chevillard2006unified,ray2008universality,chevillard2015peinture,chevillard2019skewed}, 
the dynamical origin of multiscaling from the original equations is not fully clear.
On the one hand, it has long been suggested that the presence of multifractal scaling reflects  the  spontaneous breaking of spatio-temporal symmetry of the Navier-Stokes (NS) equations and related models, which is not restored in the vanishing viscosity limit even at a statistical level \cite{kolmogorov1962refinement,frisch1995turbulence}. This qualitative viewpoint  has been recently substantiated by a series of work pointing towards the presence of a weaker (hidden) symmetry, which appears under a suitable dynamical rescaling of the space-time variables~\cite{mailybaev2021hidden,mailybaev2022hidden,mailybaev2022hiddenMT,mailybaev2023hidden}. The hidden symmetry  implies a  form of scale-invariance weaker than monofractal scaling, which  appears for example when considering ratios of increments such as Kolmogorov multipliers \cite{kolmogorov1962refinement,benzi1993intermittency,chen2003kolmogorov,eyink2003gibbsian,vladimirova2021fibonacci}.
In simplified models at least, the hidden symmetry was shown to provide a generic mechanism for spatial intermittency~\cite{mailybaev2021solvable,mailybaev2022hidden,mailybaev2022shell}, in the sense that it  fully determines the anomalous exponents of the structures functions in terms of the eigenvalues \corr of Perron-Frobenius operators acting on measures of the rescaled quantities.\rroc 

On the other hand, a different  route to intermittency is the  zero-mode mechanism, which emerges in a simpler linear turbulence problem. Namely, consider the advective-diffusive transport equation
	\begin{equation}
		\partial_t \theta+ \bu \cdot \nabla \theta = \kappa \Delta \theta+{\bf f}
	\label{eqI1}
	\end{equation}
for a passive substance $\theta(\bx, t)$, where $\bu(\bx, t)$ is a prescribed carrier flow, ${\bf f}(\bx, t)$ is a stirring force and $\kappa$ is a diffusion coefficient.
One naturally has in mind cases when the carrier flow solves the NS, a situation for  which the passive scalar is known to develop spatial intermittency in the form of ramp-cliffs structures \cite{shraiman2000scalar,iyer2018steep}. 
An instructive and  solvable framework however  emerges when one uses as a proxy for fully developed turbulence a (non-intermittent) Markovian monofractal Gaussian field: This is the framework of the Kraichnan flow theory, which by cutting  through much of the difficulties  of  turbulence theory provides  a rigorous connection between  statistical conservation laws of fluid particles and anomalous scaling of the  transported field  \cite{kraichnan1968small,chen1989probability,gawedzki1995anomalous,bernard1998slow,chen1998simulations,falkovich2001particles,frisch2007intermittency,gawedzki2008soluble}.

The purpose of the present paper is to revisit the scalar problem, in order to  highlight a connection between the two distinct mechanisms for intermittency: zero-mode and hidden symmetry.  In Eq.~\eqref{eqI1}, a technical difficulty comes from the fact that	scalar transport involves a  stochastic partial differential equation. While Kraichnan flow theory gives a meaning to  Eq.~\eqref{eqI1} for a Markovian carrier flow, we  here prefer to bypass this issue by considering a  shell-model version of Eq.~\eqref{eqI1},
\corr later precisely introduced as  the Kraichnan-Wirth-Biferale model, and where  the  multiscale nature of the scalar dynamics is  represented by a (discrete) sequence $\theta_n$ associated to a   geometric  sequence of scales  $\ell_n \propto \lambda^{-n}$. The model  is known to exhibit spatial intermittency in the form of anomalous scaling laws for the scalar structure functions, which connect to  an explicit zero-mode mechanism \cite{jensen1992shell,wirth1996anomalous,benzi1997analytic,andersen1999shell,benzi2003intermittency,biferale2007minimal}.

Despite the zero-modes, we show that the hidden symmetry approach applies to the Kraichnan-Wirth-Biferale shell model in a way similar to hydrodynamic models. It derives the multiscaling properties (existence of anomalous scaling laws, fusion rules, universality of Kolmogorov multipliers, etc.) from a single assumption, which  in spirit of the K41 theory can be interpreted as  the statistical recovery of a (hidden) scaling symmetry.  Mathematically, the recovery is formulated in terms  of a hierarchy of stationnary measures indexed by the shell numbers, connected by linear ladder operators  independent from the H\"older exponent of the velocity field.
%However, this is the hidden scaling symmetry that needs to be restored. 
%Mathematically, this means that Existence of stationnary measure of the rescaled syste,/
Our work suggests that under this hypothesis, the zero-mode solutions for correlation functions are connected to Perron-Frobenius eigenmodes, determined by the  linear operator asymptotically governing the recursion: This establishes a relation between the zero-mode and multiplicative natures of intermittency.
\rroc

%to the solution of a recursive equation specified by the linear ladder operators governing the hidden-symmetry transformation. Scaling exponents are then determined from the largest eigenvalue of the linear operator asymptotically governing the recursion. This conclusion is drawn based on qualitative analysis all quantitative results are obtained numerically.
%\MOD{We show that the hidden symmetry approach applies to the Kraichnan-Wirth-Biferale shell model in a similar way as for hydrodynamic models. It derives the multiscaling properties (existence of anomalous scaling laws, fusion rules, universality of Kolmogorov multipliers, etc.) from a single assumption in spirit of the K41 theory: the statistical recovery of scaling symmetry. However, this is the hidden scaling symmetry that needs to be restored. Mathematically, this means that Existence of stationnary measure of the rescaled syste,/Our work suggests that the zero-mode solutions for correlation functions are connected to the Perron-Frobenius eigenmodes, thereby, establishing a relation between the zero-mode and multiplicative natures of intermittency.}

The paper is organized as follows.
\S \ref{sec:Shell} presents the shell model and its general phenomenology. \S\ref{ssub:zeromode} describes the zero-mode mechanism to intermittency and its connection to statistical conservation laws. For the sake of exposition,  we here only focus on second and fourth-order structure functions, since the situation with higher order is similar but technically more sophisticated~\cite{andersen1999shell}. 
\S\ref{sec3} discusses scaling symmetries and introduces the hidden scaling symmetry as a non-linear scale invariance for  a suitably defined non-linear rescaled system. \S\ref{seq:Perron} describes the Perron-Frobenius scenario to anomalous scaling laws, which we derive  from the  hypothesis of  statistical restoring of hidden scaling symmetry. \S\ref{sec:6} highlights the implications of the Perron-Frobenius \corr scenario for the so-called fusion rules and the multifractal formalism. \S\ref{secZMvcHS} discusses connections between the zero-mode theory and the hidden symmetry approach.\rroc 
\S\ref{sec:conclusion} sketches some  concluding remarks.
In order to facilitate navigation throughout the manuscript,  some technicalities are pushed to the Appendices~\ref{app:0}--\ref{app:D}.

%%%%%%%%%%%%%%%%%%%%%%%%%%%%%%%%%
\section{Shell model for random advection}
\label{sec:Shell}

The rationale beneath random shell models for a passive scalar~\cite{wirth1996anomalous,benzi1997analytic,andersen1999shell} follows the usual  interpretation of shell models of turbulence \cite{frisch1995turbulence,biferale2003shell}, in which a multiscale dynamics is represented with a geometric sequence of scales $\ell_n = \ell_0\lambda^{-n}$. Here, $n \in \mathbb{Z}$ is an integer shell number and $\lambda > 1$ is a fixed intershell ratio usually taken equal to $2$ or fractional powers of $2$. Corresponding wavenumbers are introduced as $k_n = 1/\ell_n = k_0\lambda^{n}$. At each scale, increments of a velocity field are modeled with a shell velocity $u_n(t) \in \mathbb{R}$ and increments of a scalar field by a shell scalar $\theta_n(t) \in \mathbb{R}$; both are real variables depending on time $t$. A minimal shell model accounting for  passive scalar transport can be formulated as~\cite{jensen1992shell,benzi1997analytic,biferale2007minimal}
	\begin{equation}
	\left(\frac{d}{dt}+\kappa k_n^2\right) \theta_n = k_n\theta_{n+1}u_n-k_{n-1}\theta_{n-1}u_{n-1}, 
	\label{eq1}
	\end{equation}
with $\kappa k_n^2$ representing  diffusion term, involving the  diffusion coefficient $\kappa \ge 0$.
The right-hand side of Eq.~(\ref{eq1}) is a bilinear advection term restricted to nearest neighbors. It is designed such that the unforced ideal system (with $\kappa = 0$) conserves the scalar energy $E_\theta = \sum \theta_n^2$; see \S\ref{subsec_energy} below.

We consider constant boundary conditions at shells $n \le 0$ and $n > n_{\max}$ as
	\begin{equation}
	\theta_n(t)  \equiv \left\{\begin{array}{ll}
	1, & n \le 0; \\[2pt]
	0, & n > n_{\max}.
	\end{array} \right. 
 	\label{eq4}
	\end{equation}
The unit value at shell $n =0$ plays the role of constant large-scale forcing. The shell-number cutoff $n_{\max}$ plays the role of a Galerkin truncation, which is convenient to avoid  technical  difficulties and necessary for numerics. Because interactions are restricted to neighboring shells, setting the shells $n<0$ to a constant (e.g., unit) value is a conventional choice that does not affect the dynamics,
but later proves convenient for  the exposition of  hidden symmetry.
Let us also point out that the theory is essentially limited only by the form of symmetries and conservation laws; see~\cite{mailybaev2022hidden}. Hence, the restriction to real shell variables as well as to the so-called dyadic or Desnyansky-Novikov nonlinearity in the shell model (\ref{eq1}) is chosen for the simplicity of representation, but is not a limitation of our theory. 

\subsection{Random scalar advection}

The idea of Kraichnan~\cite{kraichnan1968small,falkovich2001particles,frisch2007intermittency,gawedzki2008soluble} is to consider the advection Eq.~(\ref{eqI1}) with  random Gaussian velocity fields, white in time and scale-invariant in space. Similarly, in the shell model (\ref{eq1}), we consider random, Gaussian shell velocities $u_n(t)$ with the correlation functions
	\begin{equation}
	\left\langle u_n(t) u_{n'}(t') \right\rangle = U_0^2 \tau_0 \, \lambda^{-n\xi } \delta(t-t') \delta_{nn'}.
 	\label{eq2}
	\end{equation}
The factor $\lambda^{-n\xi}$ mimics a rough (H\"older continuous) scale-invariant velocity field with the roughness parameter $0 < \xi < 2$. 
The  transport rate  is controlled by the typical velocity $U_0$ and associated timescale $\tau_0 = \ell_0/U_0$. Now, in order to transform Eq. \eqref{eq1} into  a SDE, one replaces the advecting velocities in Eq.~(\ref{eq1}) as 
	\begin{equation}
	u_n(t)\, dt \ \mapsto \  U_0 \tau_0^{1/2}\,\lambda^{-n\xi /2}d w_n(t),
 	\label{eq3}
	\end{equation}
where $w_n(t)$, $n \in \mathbb{Z}$, are independent real Wiener processes. Expressing also $k_n = k_0 \lambda^n$, elementary manipulations reduce Eq.~(\ref{eq1}) to the form~\cite{wirth1996anomalous}
	\begin{equation}
	\label{eq:def_strato0}
	\begin{split}
	& d \theta_n  =  \tau_0^{-1/2} \left(\gamma^n \theta_{n+1}\circ dw_n-\gamma^{n-1}\theta_{n-1}\circ dw_{n-1}\right)-\kappa k_n^2 \theta_n\, dt, 	
\end{split}
	\end{equation}
involving 	$\gamma = \lambda^{1-\xi/2}$ as the relevant velocity scaling parameter from the interval $1 < \gamma < \lambda$. \corr This equation is written using the Stratonovich convention -- see also Appendix \ref{app:0}.\rroc
%\be 
%\gamma= \lambda^{1-\xi/2} \in ]1,\lambda[.
%\ee
With the rescaling $w_n \to   w_n/ \tau_0^{1/2} $, $t \to  t/\tau_0$, %and using the relation $\tau_0 = (k_0U_0)^{-1}$, 
Eq.~(\ref{eq:def_strato0}) takes  the dimensionless form 	
\begin{equation}
	\label{eq:def_strato}
	d \theta_n  = \gamma^n \theta_{n+1}\circ dw_n-\gamma^{n-1}\theta_{n-1}\circ dw_{n-1}
	-\mathrm{Pe}^{-1} \,\lambda^{2n}\theta_n\, dt,
	\end{equation}
where $\Pe = U_0 \ell_0/\kappa$ is the dimensionless P\'eclet number.
\corr
We refer to Eq.~(\ref{eq:def_strato}) with the boundary conditions (\ref{eq4}) as the Kraichnan-Wirth-Biferale (KWB) shell model for random advection, following its explicit description as a minimal scalar model in \cite{biferale2007minimal}. 

\rroc
\subsection{Balance of scalar energy and the inertial interval} \label{subsec_energy}

We define the scalar energy at shell $n$ as $\theta_n^2$, and denote its average over random realizations of the velocity ensemble as 
$\av{\theta_n^2}$. 
One derives
the balance for averaged scalar energy at every shell $1\le n \le n_{\max}$ as (see \cite{wirth1996anomalous,biferale2007minimal} and Appendix~\ref{app:A})
	\begin{equation}
 	\label{eq7b}
	\dfrac{d\av{\theta_n^2}}{dt} = \Pi_{n-1}-\Pi_n-2 \mathrm{Pe}^{-1} \,\lambda^{2n} \av{\theta_n^2}.
	\end{equation}
Here $\Pi_n$ represents the energy flux from shell $n$ to $n+1$ due to advection, which is expressed as
	\begin{equation}
 	\label{eq8_Pi}
	\Pi_n = 
	\left\{ \begin{array}{ll}
	1, & n = 0; \\[2pt]
	\gamma^{2n}\av{\theta_{n}^2}-\gamma^{2n}\av{\theta_{n+1}^2}, & n = 1,\ldots,n_{\max}-1; \\[2pt]
	0 & n = n_{\max}.
	\end{array}\right.
	\end{equation}
%
%and reduces to the simple form
%$\Pi_n[\theta] =\gamma^{2n}\left(\av{\theta_{n}^2}-\av{\theta_{n+1}^2}\right)$ for the  inertial KB dynamics.
We are interested in the regime of developed turbulence corresponding to very large P\'eclet numbers. %, i.e., when both $\Pe$ and its logarithm are large.
 The phenomenological description of scalar dynamics then follows the usual interpretation of a turbulent cascade:  The scalar energy injected with rate $\Pi_0 = 1$ from  the forcing range $k_n/k_0 \sim 1$ is transferred to small diffusive scales
	\begin{equation}
	\frac{k_n}{k_0} = \lambda^n \sim \Pe^{1/\xi}, 
	\label{eq_A_1}
	\end{equation}
 where it is absorbed~\cite{frisch1995turbulence,wirth1996anomalous}.
The condition \eqref{eq_A_1} is obtained by comparing the diffusion term in Eq.~(\ref{eq7b}) with the flux term \eqref{eq8_Pi}, which yields $\gamma^{2n} \sim \mathrm{Pe}^{-1}\lambda^{2n}$ with $\gamma = \lambda^{1-\xi/2}$.  %under the condition $D_n \gtrsim \gamma^{2n}$. Using $D_n = \Pe^{-1}\lambda^{2n}$ and $\gamma = \lambda^{1-\xi/2}$, this condition is written as
%	\begin{equation}
%	\frac{k_n}{k_0} = \lambda^n \gtrsim \Pe^{1/\xi}.
%	\label{eq_A_1}
%	\end{equation}
%It defines the diffusion range of very large wavenumbers, i.e., very small scales $\ell_n = 1/k_n$. On opposite side, at large scales, we define the forcing range as $k_n/k_0 \sim 1$.

The present work will focus on describing the statistical properties within the (inertial) interval of  scales at which both the forcing and diffusion terms are negligible. The inertial interval corresponds to the wavenumbers satisfying the conditions
\be
	1 \ll \frac{k_n}{k_0} \ll \Pe^{1/\xi}.
	\label{eq:inertial}
\ee
In terms of shell numbers, it reads $0 \ll n \ll   n_K$ with the diffusive (Kolmogorov) shell number
\be
	n_K = \dfrac{1}{\xi}\log_\lambda \mathrm{Pe}.
	\label{eq:inertial_n}
\ee
Neglecting the forcing and diffusion terms in Eq.~(\ref{eq:def_strato}) yields 
\corr
	\begin{equation}
 	\label{eq5id}
		d \theta_n  
		= \gamma^n \theta_{n+1}\circ dw_n-\gamma^{n-1}\theta_{n-1}\circ dw_{n-1}, 
		\quad
		n \in \mathbb{Z},
	\end{equation}
\rroc
to which we refer as the ideal KWB model. We remark that neglecting the large-scale (forcing) and small-scale (diffusion) cutoffs implies that the ideal KWB model is considered formally for arbitrary $n \in \mathbb{Z}$.
The energy balance provided by Eqs.~(\ref{eq7b}) and (\ref{eq8_Pi}) reduces in the inertial interval to
	\begin{equation}
 	\label{eq7b_Id}
	\dfrac{d\av{\theta_n^2}}{dt} = \Pi_{n-1}-\Pi_n
	= \gamma^{2n-2}\av{\theta_{n-1}^2}-(\gamma^{2n-2}+\gamma^{2n})\av{\theta_{n}^2}+\gamma^{2n}\av{\theta_{n+1}^2}.
	\end{equation}
% The presence of the (infrared) forcing is however essential: It allows for  steady-state  solutions   with (constant) unit flux $\Pi_n=1$ -- see also \S \ref{ssub:zeromode}.

Please observe that, from a mathematical point of view, the study of inertial interval statistics is formulated as the asymptotic analysis of the KWB model \eqref{eq:def_strato}  in the limit  
	\begin{equation}
 	\label{eq7b_II}
	t \to \infty, \quad
	 n_{\max}\to \infty, \quad \Pe \to \infty, \quad n \to \infty,
	\end{equation}
where the limits are considered successively from left to right. These limits signify the large-time asymptotic (or time averaging), removal of small-scale cutoff, vanishing diffusion limit and, finally, small-scale asymptotic. 

\subsection{Numerical model}

Theoretical assumptions and conclusions of this paper are verified with numerical simulations designed to give accurate statistics in a large inertial interval. Our numerical model is given by Eq.~(\ref{eq:def_strato}) with the boundary conditions (\ref{eq4}), where we set $\lambda =2^{1/3}$, $n_{\max} = 36$ and $\Pe = \lambda^{n_K\xi} \approx 140$ with $n_K = 32$ and $\xi = 2/3$.  
We expect that the inertial interval statistics does not depend on a specific form of the diffusion mechanism, except perhaps from finite-$\Pe$ effects~\cite{wirth1996anomalous,benzi1997analytic}. \corr For our numerical analysis, it  proves convenient to remove the diffusion term completely from Eq.~(\ref{eq:def_strato}) 
for the shells  $n < n_K$ in the  inertial interval. 
With this choice, the energy balance \eqref{eq7b_Id} is exactly satisfied for the shells $n = 1,\ldots,n_K-1$.
Our numerical simulations of the KWB dynamics use the Milstein scheme \cite{higham2021introduction}, as described in Appendix~\ref{app:0}; Examples of times series are shown in the inset of Figure \ref{fig:1}(a).
\rroc

%%%%%%%%%%%%%%%%%%%%%%%%%%%%%%%%%%%%%%%%%%%%%%%
\section{Zero-mode scenario to anomalous scaling laws}
\label{ssub:zeromode}
\begin{figure}[t]
\begin{minipage}{0.49\textwidth}
\includegraphics[width=\textwidth,trim=0cm 0cm 0cm 0cm, clip]{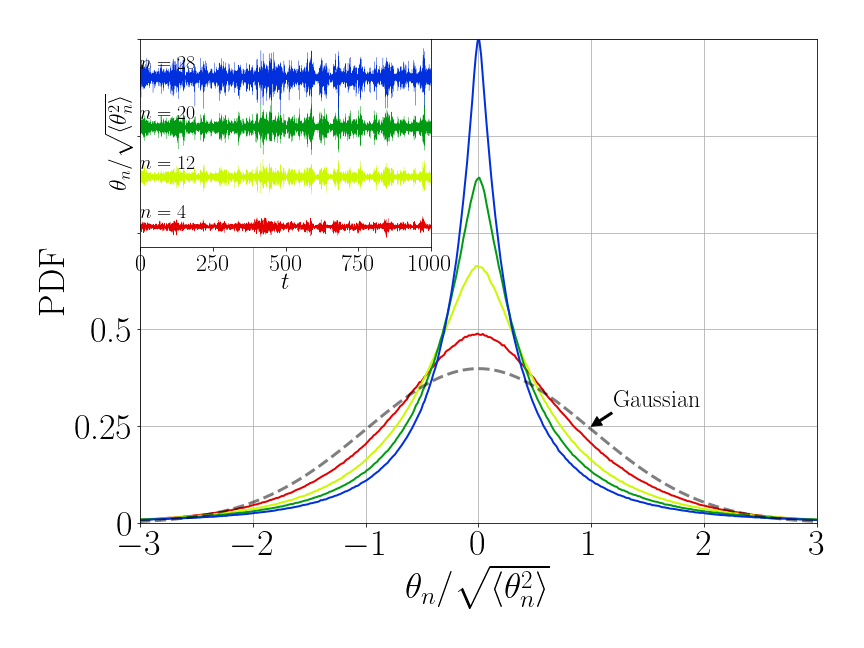}
\end{minipage}
\begin{minipage}{0.49\textwidth}
\includegraphics[width=\textwidth,trim=0cm 0cm 0cm 0cm, clip]{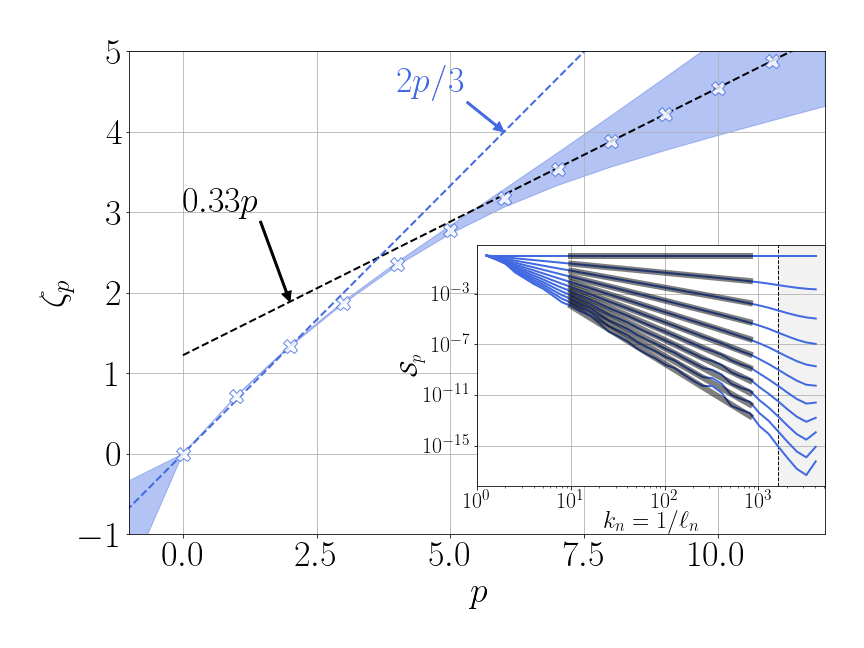}
\end{minipage}
\setlength{\unitlength}{\columnwidth}
\begin{picture}(1,0)(0,0)
\put(0.26,0.37){(a)}
\put(0.75,0.37){(b)}
\end{picture}
\caption{(a) PDF of  inertial-range shell variables normalised by their variance. Inset shows corresponding time-serie realizations.
(b) Scaling exponents $\zeta_p$. Inset shows power law scalings of structure functions for the integer orders from $p = 0$ (top) to $p=10$ (bottom), with the grey thick lines indicating the scaling region.}
\label{fig:1}
\end{figure}
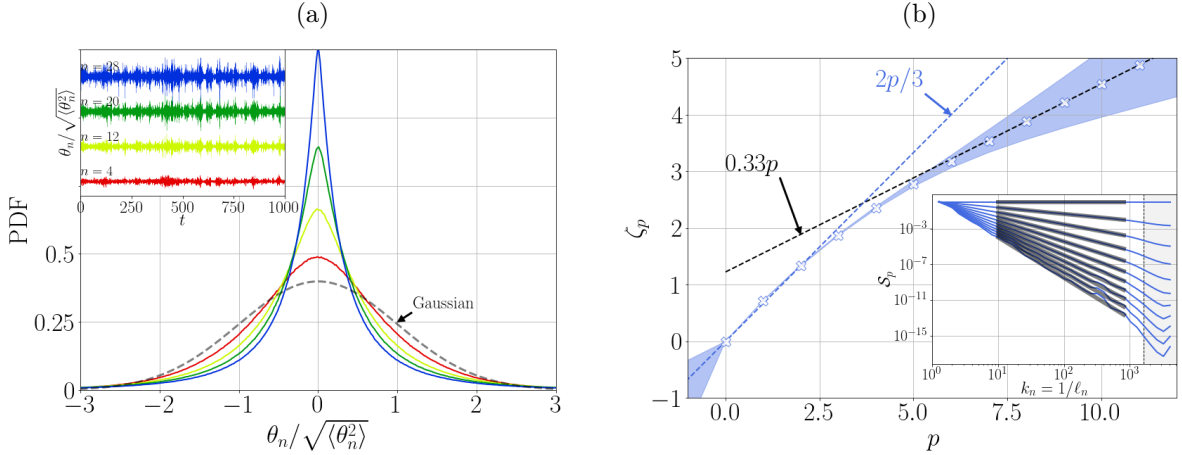

Structure functions are traditional observables for the analysis of intermittency in fully developed turbulence \cite{frisch1995turbulence}. 
In shell models, they are usually defined as the time-ensemble averaged moments
\be
	\label{eq:Sp}
	{\mathcal S}_p (\ell_n)  = \lim_{T\to \infty}T^{-1}\int_0^ T \langle |\theta_n|^p \rangle \,dt.
\ee
In the inertial range, the KWB model  is known to exhibit power-law scaling laws  
\corr
\be
	\label{eq:SpSL}
	{\mathcal S}_p (\ell_n) \propto \ell_n^{\zeta_p} \propto \lambda^{-\zeta_pn},\quad p >-1
\ee
\rroc
with non-linear (anomalous) behavior of the exponents $\zeta_p$. This  distinctive feature  reflects the lack of  statistical scale-invariance for the distributions of  the shell variables  \cite{frisch1995turbulence} -- See Fig.~\ref{fig:1}.
For even-order structure functions, the presence of anomalous exponents can be explicitly related to a zero-mode mechanism.
Following \cite{benzi1997analytic, andersen1999shell}, \corr this  section briefly  accounts for the zero-mode theory at the level of second and fourth-order correlators. \rroc  %We also point out that the presence of zero-modes reflects the presence of exact statistical conservations laws in the ideal model.

\subsection{Second-order structure functions}\label{subs_flux}

% We now consider a statistically stationary state and write the second-order correlator $S_2(k_n)$.
Performing the time-average of  the energy balance in the  inertial interval, namely Eq.~(\ref{eq7b_Id}),
the left-hand side vanishes. This implies  in particular that the mean flux of scalar energy is  constant in the inertial interval,
prescribed  by injection condition $\Pi_0 = 1$, consistent with the phenomenology of \S \ref{subsec_energy}. Besides, replacing $\av{\theta_n^2}$ by their time averages from Eq.~(\ref{eq:Sp}), and cancelling the common factor $\gamma^{2n-2}$ in Eq.~(\ref{eq7b_Id}) yields the recursion
	\begin{equation}
 	\label{eq7B1}
	{\mathcal S}_2(\ell_{n-1})-(1+\gamma^{2}) \mS_2(\ell_{n}) +\gamma^{2}S_2(\ell_{n+1}) = 0.
	\end{equation}

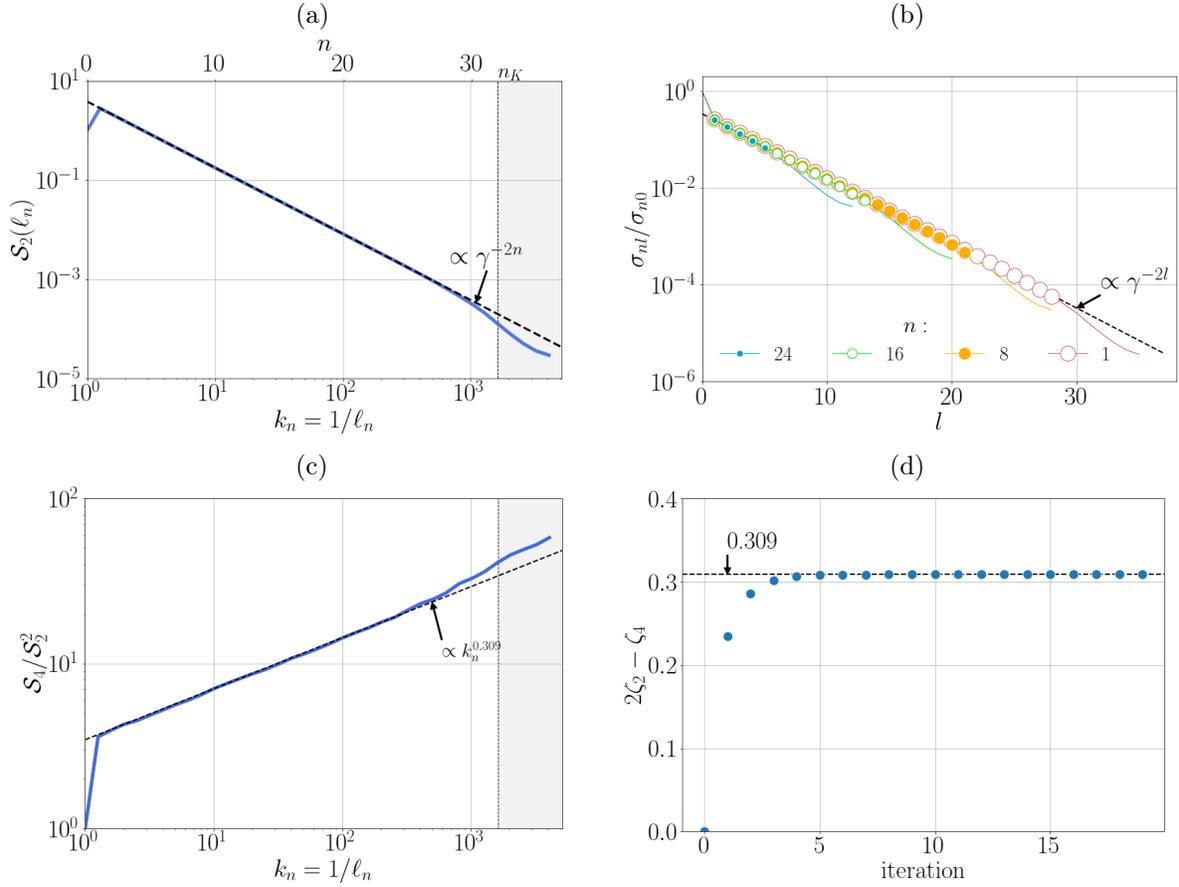
\begin{figure}
\includegraphics[width=0.49\textwidth,trim=0cm 0cm 0cm 0cm, clip]{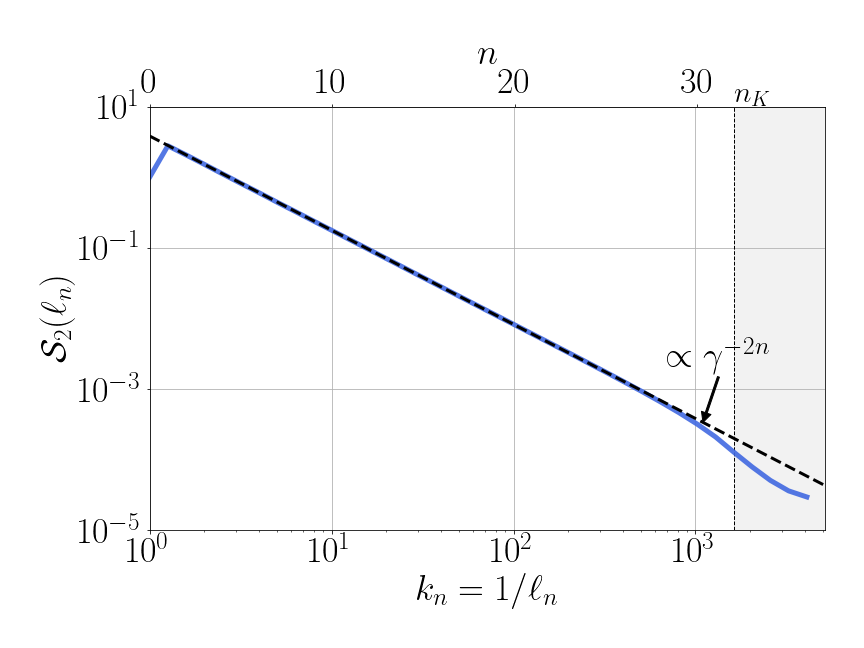}
\includegraphics[width=0.49\textwidth,trim=0cm 0cm 0cm 0cm, clip]{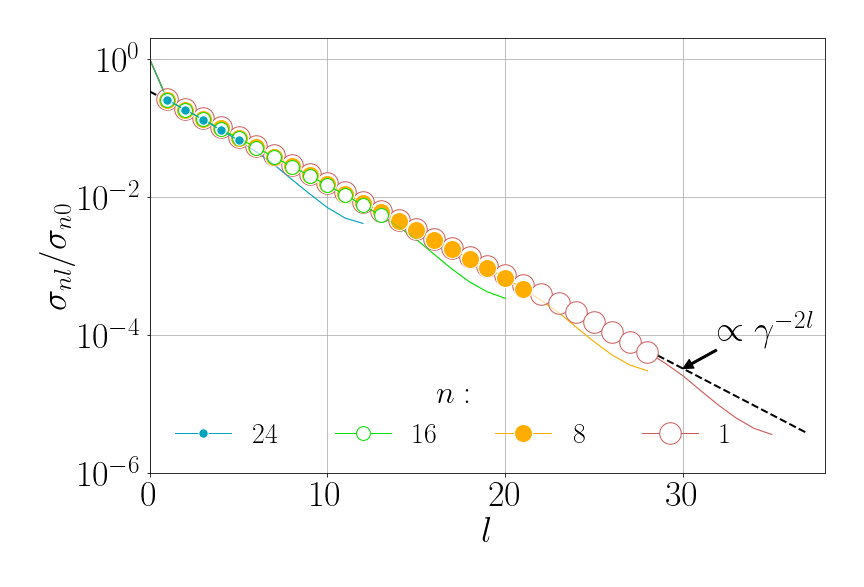}\\
\includegraphics[width=0.48\textwidth,trim=0cm 0cm 0cm 0cm, clip]{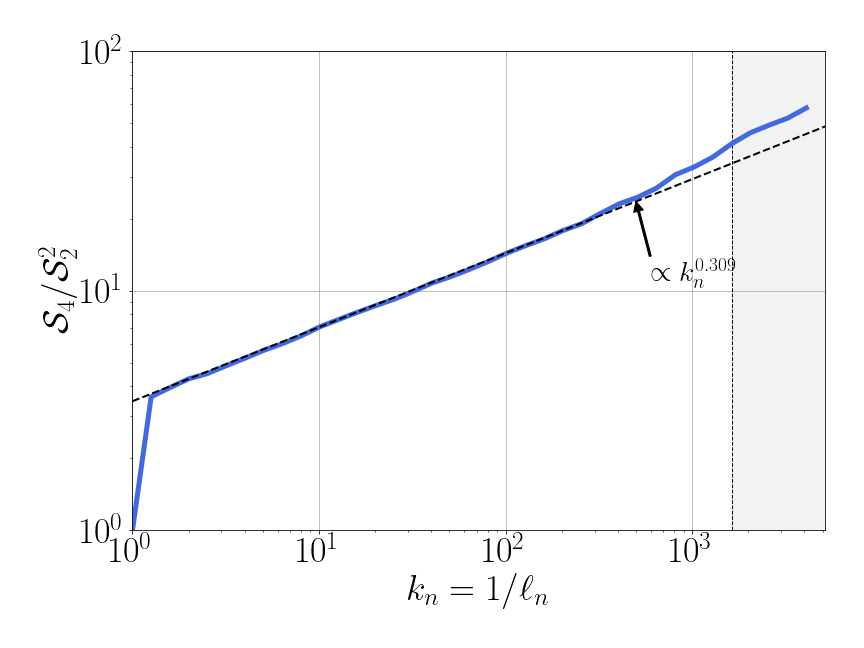}
\includegraphics[width=0.48\textwidth,trim=0cm 0cm 0cm 0cm, clip]{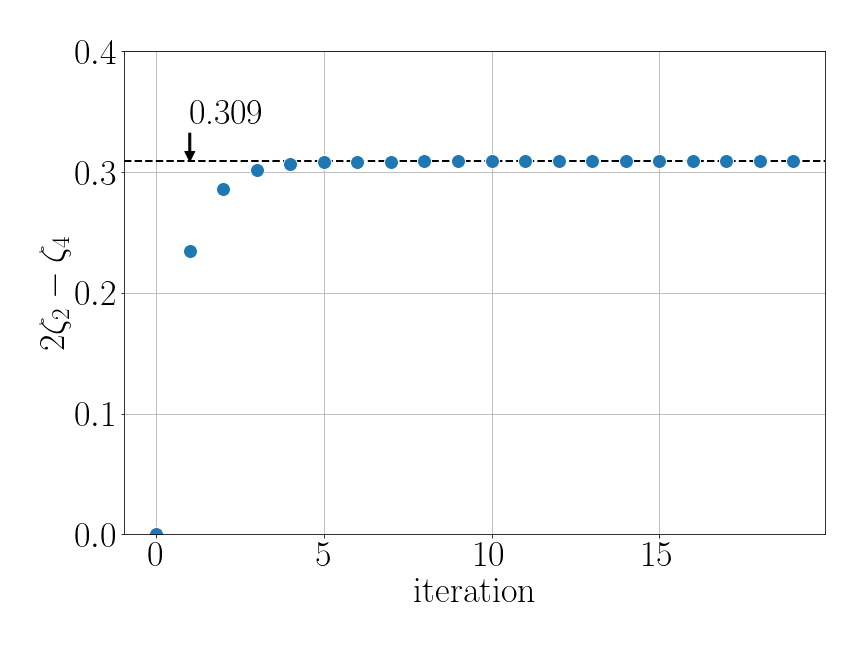}
\setlength{\unitlength}{\columnwidth}
\begin{picture}(1,0)(0,0)
\put(0.26,0.755){(a)}
\put(0.74,0.755){(b)}
\put(0.26,0.39){(c)}
\put(0.74,0.39){(d)}
\end{picture}
\caption{(a) Second order structure function with the Kolmogorov scaling $S_2(\ell_n) = \ell_n^{\zeta_2}$. (b) Collapse of off-diagonal terms $\sigma_{nl}/\sigma_{n0}$. Dots indicate  the inertial interval $1\le n,n+l \le 28$. (c) Flatness with the zero-mode scaling exponent $2\zeta_2-\zeta_4 \simeq 0.309$. (d) Convergence of the iteration procedure  to the zero-mode scaling exponent; \corr see Appendix~\ref{app_ZM4}.\rroc }
\label{fig:2}
\end{figure}

\corr
Requiring $S_2(\ell_n) \to 0 $ for large $n$,
the recursion yields the looked-for 
 power-law solutions  \eqref{eq:SpSL}, explicitly ${\mathcal S}_2 (\ell_n) \propto \lambda_2^n$ with $\lambda_2 = \lambda^{-\zeta_2}$  a root of the second-order polynomial
	\begin{equation}
 	\label{eq7B1_P}
	\gamma^{2}\lambda_2^2-(1+\gamma^{2}) \lambda_2 +1 = 0.
	\end{equation}
We rule out the root $\lambda_2 = 1$, for which $\mS_2(\ell_{n})  \propto 1$ does not decrease at small scales. The remaining solution $\lambda_2 = \gamma^{-2} = \lambda^{\xi-2}$ yields
\rroc	
the scaling law~\cite{wirth1996anomalous}
	\begin{equation}
 	\label{eq7B3_v1}
	\mS_2(\ell_{n})  \propto \ell_n^{\zeta_2} \ \ \text{with} \ \
%	\end{equation}
%with the relation
%	\begin{equation}
% 	\label{eq7B3r}
	\zeta_2 =  2-\xi.
	\end{equation}
The exponent $\zeta_2$ provides a non-vanishing energy flux (\ref{eq8_Pi}) from large to small scales through the inertial interval. This property is analogous to the relation of the third-order structure function with the energy flux in turbulence~\cite{frisch1995turbulence}. In our numerics,  the scaling  \eqref{eq7B3_v1} is indeed observed over almost three decades -- see Fig. \ref{fig:2}(a).

\subsection{Fourth-order structure functions}
\label{subsec_S4}

In the KWB model, the scaling exponents associated to higher-order even correlators \eqref{eq:Sp} are determined by solving zero-mode \corr (linear homogeneous) systems leading to eigenvalue problems similar to \eqref{eq7B1_P}: This is the zero-mode mechanism for intermittency.  To explicitly describe the case that gives non-Kolmogorov (anomalous) scaling, we now consider fourth-order structure functions.
Let us extend definition \eqref{eq:Sp} into
\be
	\sigma_{nl} = \lim_{T\to \infty}T^{-1} \int_0^T 
	\av{\theta_{n}^2\theta_{n+l}^2} dt,
	\label{eq:4order}
\ee
which involve two different shells $n$ and $n+l$ from the inertial interval.
% Clearly, one has $S_4(\ell_n) = \sigma_{n0}$.  
For these structure functions, one derives the following system of equations (see Appendix \ref{app:A})
\be
-a_l \sigma_{nl}  + b_l \sigma_{n(l+1)}+  b_l\gamma^{-2} \sigma_{n(l-1)}
+  b_{-l} \sigma_{(n+1)(l-1)}+  b_{-l}\gamma^{-2} \sigma_{(n-1)(l+1)} = 0
\label{eq:ds4B}
\ee
with the coefficients
\begin{equation}
 	\label{eqZ3}
	a_l = \gamma^{l}+\gamma^{-l-2} + \gamma^{-l}+\gamma^{l-2} + 4\gamma^{-l} \delta_{l1}, \quad
	b_l = \gamma^{l}+2\delta_{l0}.
\end{equation}
This is the zero-mode system with respect to $\sigma_{nl}$.
\rroc

Following~\cite{l1996towards,benzi1997analytic,biferale2003shell}, a proper zero-mode solution is defined by the combined use of \emph{(i)} a power-law scaling ansatz for the original structure function
 	\begin{equation}
 	\label{eq7B3}
	S_4(\ell_n) = \sigma_{n0} \propto \ell_n^{\zeta_4}
	\end{equation}
with the unknown exponent $\zeta_4$, and \emph{(ii)} the so-called fusion rule ansatz
	\be
 	\label{eqZ6}
	%S_4(k_n) = \lambda^{-\zeta_4}  S_4(k_{n-1})  , \quad
	\sigma_{nl} = C_{l}\sigma_{n0}
	\ee
with the unknown coefficients $C_l$. 
\corr
Clearly, $C_0 = 1$. 
These coefficients must behave as \rroc
 	\begin{equation} 	
	\label{eqZM14}
	C_l \to 0 \textrm{ \ as \ } l \to \infty,
	\end{equation}
which requires that moments $\sigma_{nl}$ decrease for large $l$.

The solution proceeds as follows.
Using relations  \eqref{eq7B3} and \eqref{eqZ6}, one expresses every term in Eq.~(\ref{eq:ds4B}) as a multiple of $\sigma_{n0}$. For example,
	\be
	\sigma_{(n+1)(l-1)} =  C_{l-1}\sigma_{(n+1)0} = C_{l-1}\lambda^{-\zeta_4} \sigma_{n0}.
	\label{eq_P_deriv}
	\ee
\corr
In the case $l = 0$, one should also use the symmetry property following from Eq.~(\ref{eq:4order}), e.g., $\sigma_{n(-1)} = \sigma_{(n-1)1}$.
In this way, after dividing by $\lambda^{\zeta_4}\sigma_{n0}$, one reduces Eq.~(\ref{eq:ds4B})  to the form
	\be
	\left\lbrace
	\begin{split}
	& 
	(-a_0 C_0
	+2b_0 C_{1} )\lambda^{-\zeta_4}
	+2b_0 \gamma^{-2} C_{1}
	 = 0, 
	\quad
	l =0,\\
	& 
	b_{-l}  \lambda^{-2\zeta_4} C_{l-1} 
	+(-a_l C_l  + b_l C_{l+1} +  b_l\gamma^{-2} C_{l-1})\lambda^{-\zeta_4}  
	+b_{-l}\gamma^{-2} C_{l+1}
	 = 0, \quad
	l \ge 1.
	\end{split}
	\right.
	\label{eq:ds4BX}
	\ee	
Denoting $\lambda_4 = \lambda^{-\zeta_4}$, one can write the system (\ref{eq:ds4BX}) as the second-order eigenvalue problem
\be
 	\label{eqZM_EP4}
	(\lambda_4^2 M + \lambda_4 K + L)C = 0, 
\ee
where $C = (C_0,C_1,\ldots)$ is the infinite-dimensional eigenvector normalized with the first component $C_0 = 1$. The infinite-dimensional matrices $M$, $K$ and $L$, whose elements one can infer from Eq.~(\ref{eq:ds4BX}), have one- or three-diagonal structrue.

The eigenvalue problem (\ref{eqZM_EP4}) reminds Eq.~(\ref{eq7B1_P}) for the second-order structure functions. Its three-diagonal structure allows to find solutions in a semi-analytic way as described in \cite{benzi1997analytic}; See Appendix~\ref{app_ZM4} for a slightly modified form of this derivation. In our numerics, taking $\xi=2/3$ we obtain $\zeta_4 \simeq 2.357$. This exponent deviates from the monofractal prediction, i.e., $\zeta_4 \ne 2\zeta_2$. The deviation $\zeta_4-2\zeta_2 \simeq -0.309$ provides the scaling exponent for the flatness $S_2(\ell_n)/S_2^2(\ell_n) \propto \ell_n^{\zeta_4-2\zeta_2}$, which increases at small scales.
As shown in Fig. \ref{fig:2}(c), the anomalous scaling for the flatness agrees with the zero-mode prediction over almost the full three decades of our numerics, with the  deviations seen at the ultraviolet end being caused by finite-size effects.
Besides,  our numerics support the fusion rule ansatz  \eqref{eqZ6}. Figure \ref{fig:2}(b) shows that the coefficients 
$\sigma_{nl}/\sigma_{n0}$ become indeed independent of the index $n$ when both $n$ and $ n+l$ lie in inertial range, and feature the Kolmogorov scaling $C_l \propto \gamma^{-2l}$.
\rroc

%##############################################
%HIDDEN SYMMETRY
\section{Classical vs Hidden scaling symmetries}
\label{sec3}
\corr %
We now present the approach on multiscaling based on the concept of hidden scaling symmetry. 
This symmetry emerges under a proper rescaling of the ideal model \eqref{eq5id}. 
%This rescaling can be seen as a projection in phase space, which maps original variables to their ratios. 
We argue and verify numerically that the hidden symmetry (HS)  is restored in the statistical sense for scales in the inertial interval, even though the original scaling symmetries get broken. The anomalous power-law scaling and fusion rules follow from the statistically restored hidden symmetry, thereby, providing a first-principle route to the multifractality. The HS-approach is qualitative,  aiming to understanding statistics in the inertial interval in greater details, rather than computing specific exponents accessible by the zero-mode theory. We postpone further comparison with the zero-mode approach to Section~\ref{secZMvcHS}, and focus now on the hidden symmetry and its implications.
\rroc

\subsection{Classical  scaling symmetries}
\label{sec3A}
Let us first recall that the ideal model (\ref{eq5id}) possesses a family of scaling symmetries generated by the transformations
	\begin{equation}
	\label{eqS1a}
	% n \mapsto n+1, \quad 
	\theta_n \mapsto \lambda^h \theta_{n\pm1}, \quad 
	w_n \mapsto \gamma^{\pm1} w_{n\pm1},\quad 
	t \mapsto  \gamma^{\pm2} t.
	\end{equation}
The change of shell index $n \mapsto n+1$ (and similarly $n \mapsto n-1$) defines the spatial scaling by the factor of $\lambda$, since $\ell_{n+1} = \ell_n/\lambda$, while the exponent $h \in \mathbb{R}$ defines rescaling of scalar variables by an arbitrary factor $\lambda^h$.
The change of time $t' = \gamma^2 t$ combined with the scaling $w'_n = \gamma w_{n+1}$ yield $w'_n(t') = \gamma w_{n+1}(t'/\gamma^2)$, which is again a Wiener process \cite{evans2012introduction}. While  the (full) KWB dynamics  \eqref{eq:def_strato} breaks the  symmetries (\ref{eqS1a}) at large and small scales due to the boundary conditions and diffusion, classical turbulence theory \cite{frisch1995turbulence} postulates that these symmetries may be recovered in a statistical sense in the inertial interval, i.e., in the limits prescribed by  Eq.~\eqref{eq7b_II}. Because the family of symmetries (\ref{eqS1a}) contains an arbitrary exponent $h$, one expects that the stationary measure exhibits a statistical intermittency in the form of multifractal scaling~\cite{frisch1980fully}. Heuristically, this means that the exponent  $h$ is not selected univocally but rather as a probability distribution. The idea of hidden symmetry is that this degeneracy can be removed by a suitable rescaling (projection) of the phase space and time. The present section describes the construction for the KWB  model.

\subsection{Shell-time rescaling}
\label{sec3B}
Given a sequence of scalars $\theta = (\theta_n)_{n \in \mathbb Z}$, we introduce a sequence of scalar amplitudes $\mathcal{A}[\theta] = (\mathcal{A}_n[\theta])_{n \in \mathbb Z}$ as
	\be
 	\label{eqS2}
	\mathcal{A}_n[\theta] = \sqrt{\theta_{n}^2+\alpha \theta_{n-1}^2+\alpha^2 \theta_{n-2}^2+\cdots},
	\ee
where $\alpha$ is a fixed positive pre-factor. 
The amplitudes $\mathcal{A}_n[\theta]$ are positive and represent local averages of the scalar $\theta$ about the scale $\ell_n$.
Later in \S \ref{sec:multifractal} we justify that the locality is ensured under the condition $\alpha < \lambda^{-2h_{\max}}$ with $h_{\max} \approx 1.4$. 

Let us now fix a reference shell $m \ge 0$ and define the respective rescaled variables. Using the scalar amplitudes \eqref{eqS2}, we define the rescaled scalar   $\Theta^{(m)} = \big(\Theta^{(m)}_N\big)_{N \in \mathbb{Z}}$, the rescaled Wiener process  $W^{(m)} = \big(W_N^{(m)} \big)_{N \in \mathbb{Z}}$  and proper time $\tau^{(m)}$ as
	\begin{equation}
 	\label{eqS3}
	\Theta_N^{(m)} = \frac{\theta_{m+N}}{\mathcal{A}_m[\theta]}, \quad 
	W_N^{(m)} = \gamma^m w_{m+N}, \quad 
	\tau^{(m)} = \gamma^{2m} t.
	\end{equation}
We note that $\big(W_N^{(m)} \big)_{N \in \mathbb{Z}}$ is still a sequence of independent Wiener processes in terms of the rescaled time $\tau^{(m)}$. For $m = 0$, the boundary conditions \eqref{eq4} yield $\mathcal{A}_0[\theta] =1/\sqrt{1-\alpha}$, and Eq.~\eqref{eqS3} prescribes a linear rescaling. For $m >0 $, the transformation (\ref{eqS3}) is nonlinear. Using Eqs.~\eqref{eqS2} and \eqref{eqS3}, one can verify the  identity, valid for all $m \ge 0$:
	\begin{equation}
 	\label{eqS3b}
	 	\mathcal{A}_0[\Theta^{(m)}] \equiv 1.
	\end{equation}

We remark that the specific form of amplitudes (\ref{eqS2}) is not crucial. Essentially, one needs the scale-invariance, homogeneity
 and positivity  of $\mathcal{A}_n[\theta]$.
 We refer to \cite{mailybaev2022hidden} for a general theory that relates this ambiguity to equivalent representations of projections in phase space. Such algebraic interpretation applies to our model too, but goes beyond our present scope.
%we will not address it here. For simplicity of derivations, we restrict our analysis to definition (\ref{eqS2}).

\subsection{Hidden scaling symmetry}
\label{sec3C}

We now consider both the reference shell $m$ and shells $n = m+N$ of rescaled variables from the inertial interval, where the dynamics is governed by the ideal model (\ref{eq5id}).
\corr Using the new variables $\Theta^{(m)}, W^{(m)}, \tau^{(m)}$ from Eq. \eqref{eqS3},  
the ideal model rescales into the Hidden KWB form (see Appendix \ref{app:B})
\be
	\label{eq:rescaled}
	\begin{array}{rl}
	d\Theta_N
	= & \displaystyle 
	\gamma^N \Theta_{N+1}\circ dW_N-\gamma^{N-1}\Theta_{N-1}\circ dW_{N-1} \\[3pt]
	&-\Theta_N \sum_{J\le0} \alpha^{-J}\Theta_J \left(
	\gamma^J \Theta_{J+1}\circ dW_J-\gamma^{J-1}\Theta_{J-1}\circ dW_{J-1}
	\right),
 	\quad N\in \mathbb{Z}.
	\end{array}
\ee
Here we have (temporarily) omitted the superscript $(m)$ to emphasize the independence of the rescaled system from the reference shell $m$. Such independence means that 
the  change of $m$ generates a symmetry of the rescaled equations (\ref{eq:rescaled}).
\rroc

This symmetry can be introduced in terms of positive and negative unit shifts acting on the scalar and Wiener sequences and time as
    \begin{equation}
	 m\mapsto m \pm 1: \hspace{1cm}  \Theta  \mapsto  a^{\pm 1}[\Theta], \quad 
	W \mapsto b^{\pm 1}[ W],\quad 
	\tau \mapsto  \gamma^{\pm 2} \tau.
	\label{eq:HS}
    \end{equation}
The ladder generators $a^{\pm 1}$ and $b^{\pm 1}$ 
are derived by comparing the processes $\Theta^{(m)}(\tau^{(m)})$ and $W^{(m)}(\tau^{(m)})$ for adjacent $m$ as shown Appendix~\ref{app:C}. In terms of components, they read
    \begin{equation}
	a^{+1}_N[\Theta] = \dfrac{\Theta_{N+1}}{\sqrt{\alpha+\Theta_1^2}}, \quad
	a^{-1}_N [\Theta] = \sqrt{\dfrac{\alpha}{1-\Theta_0^2}}\, \Theta_{N-1},\quad
	b^{\pm 1}_N [W] = \gamma^{\pm 1}W_{N\pm 1}.
	\label{eq:apm}
    \end{equation}
The $m\mapsto m+1$ and $m\mapsto m-1$ transformations (\ref{eq:HS}) are inverse to each other, and they leave the rescaled system (\ref{eq:rescaled}) invariant. This can be checked from direct calculation, but is rather a direct consequence of the construction: The transformations \eqref{eq:HS}
 map the rescaled processes $\Theta^{(m)}(\tau^{(m)})$ and $W^{(m)}(\tau^{(m)})$ into $\Theta^{(m\pm1)}(\tau^{(m\pm1)})$ and $W^{(m\pm1)}(\tau^{(m\pm1)})$, which in turn  satisfy the ($m$-independent) rescaled system \eqref{eq:rescaled}.

We call the invariance of  the rescaled system with respect  to transformations  \eqref{eq:HS} and \eqref{eq:apm} the \emph{hidden scaling symmetry}, in analogy with the hidden symmetries  obtained earlier for different turbulence models~\cite{mailybaev2021hidden,mailybaev2022hidden,mailybaev2022hiddenMT}. 
Compositions of transformations (\ref{eq:HS}) generate a group of symmetries given by operators $a^J$ and $b^J$ for any $J \in \mathbb{Z}$ together with the time rescaling $\tau \mapsto  \gamma^{2J} \tau$, which correspond to the shift $m \mapsto m+J$ of the reference shell.  In particular, these transformations map the rescaled process $\Theta^{(m)}(\tau^{(m)})$ into $\Theta^{(m+J)}(\tau^{(m+J)})$. 
At a heuristic level, the hidden symmetry can be interpreted as a fusion of the classical scaling symmetries  \eqref{eqS1a} depending on the H\"older exponent $h$ into the $h$-independent symmetry (\ref{eq:HS}). This yields a new and weaker form of symmetry, which may be restored in a solution even when all classical scaling symmetries are broken. In particular, this refers to a statistically stationary state as we describe in the following subsection.

%Finally, we point out that compositions of transformations (\ref{eq:HS}) generate a group of symmetries given by operators $a^J$ and $b^J$ for any $J \in \mathbb{Z}$ together with the time rescaling $\tau \mapsto  \gamma^{2J} \tau$, which correspond to the shift $m \mapsto m+J$ of the reference shell.  In particular, these transformations map the rescaled process $\Theta^{(m)}(\tau^{(m)})$ into $\Theta^{(m+J)}(\tau^{(m+J)})$. 

\begin{comment}
\subsection{Statistical hidden symmetry}
The hidden symmetry can also be considered in the statistical sense: Let $\mu^\infty$ be a stationary (time-averaged) probability measure associated to a statistical solution $\Theta$ of the ($m$-independent)  rescaled system \eqref{eq:rescaled}- \eqref{eq:rescaled_B}, which is formally given by 
    \be
    \mu^{\infty}(\mathfrak{DX})
    = \lim_{T \to \infty} T^{-1} \int_0^T
	d\tau\,\av{\mathbbm{1}_{\Theta (\tau) \in \, \mathfrak{DX}}}
	\label{eq_SHS1}
    \ee
for any measurable subset $\mathfrak{DX}$ in the phase space $\Theta = (\Theta_N)_{N \in \mathbb{Z}}$.
In particular, the invariant measure in the set of semi-infinite sequence  $\Theta = $, is defined through the finite-dimensional marginals
\be
	\mu^\infty(dx_0,dx_{-1},dx_{-2}\cdot dx_{-k})=    \lim_{T \to \infty} T^{-1} \int_0^T
	d\tau\,\av{\mathbbm{1}_{\cup \Theta_i (\tau) \in \, \mathfrak{dx_0....}}}
\ee
\end{comment}
\subsection{Statistically restored hidden symmetry}
\label{sec3D}

Recall that the mathematical formulation of the inertial interval corresponds to the limit (\ref{eq7b_II}). For the reference shell $m$, it takes the form
	\begin{equation}
 	\label{eq:limits}
	t \to \infty, \quad
	 n_{\max}\to \infty, \quad \Pe \to \infty, \quad m \to \infty.
	\end{equation}
In this limit, the rescaled ideal system (\ref{eq:rescaled}) is valid.  
We now formulate the hidden scaling symmetry of the rescaled system in the statistical sense.
To that end, we adopt the ergodicity assumption, which prescribe the physically relevant measure in terms of  time averages as 
    \be
   \mathbb{P}^{(m)}(\mathfrak{A})
    = \lim_{T \to \infty} T^{-1} \int_0^T
	d\tau^{(m)}\,\av{\mathbbm{1}_{\Theta^{(m)}(\tau^{(m)}) \in \, \mathfrak{A}}}
	\label{eq_SHS1}
    \ee
for any measurable subset $\mathfrak{A}$ in the phase space $\Theta = (\Theta_N)_{N \in \mathbb{Z}}$.
%Please note, that  the average in Eq.~(\ref{eq_SHS1}) is defined with respect to the rescaled time $\tau^{(m)}$.
  %We now adopt the ergodicity assumption, which expresses time averages of observables $\varphi[\Theta]$ in terms of measure averages as
%   	\be
  % 	\lim_{T \to \infty} T^{-1} \int_0^T
%	\varphi\left[ \Theta^{(m)}(\tau^{(m)}) \right] d\tau^{(m)}
%	= \mathbb{E}^{(m)}_{\Theta} \left( \varphi[\Theta] \right),
%	\label{eq_SHS1erg}
 %   	\ee
The  averages of observables $\varphi[\Theta]$ then identify to their expectation values with respect to $\mathbb{P}^{(m)}$, i.e.
%where 
   	\be
	\mathbb{E}^{(m)} \left( \varphi \right)
	= \int \varphi[\Theta] \, \mathbb{P}^{(m)}(d \Theta).
	\label{eq_SHS1exp}
    	\ee
%is the expectation value with respect to $\mathbb{P}^{(m)}$.

The hidden symmetry  transformation (\ref{eq:HS}) performs the shift $m \to m\pm 1$. 
The corresponding time change $\tau \mapsto \gamma^{\pm 2}\tau $ is linear and does not affect the defining time average in Eq.~\eqref{eq_SHS1}. % and, therefore, does not affect the statistically stationary state.  
Hence,  the shift   relates  the measures associated to adjacent reference shells through the pushforward transformations $a^{\pm 1}_\sharp$ as
    \be
    \mathbb{P}^{(m\pm 1)} = a^{\pm 1}_ \sharp \mathbb{P}^{(m)}.
	\label{eq_SHS1b}
    \ee
%\footnote{Please observe,  that the property \eqref{eq_SHS1b} comes from the linear nature of the  time change $\tau^{(m)} \to \tau^{(m+1)}$, prescribed by  Eq.~(\ref{eq:HS}). It is different from the more sophisticated situation in the  turbulence models studied in~\cite{mailybaev2021hidden,mailybaev2022hidden}, where the time rescaling is also nonlinear.}
We say that the hidden symmetry is restored statistically in the inertial interval, if the measures $\mathbb{P}^{(m)}$ restricted to the inertial-interval variables are invariant under the hidden symmetry transformation (\ref{eq_SHS1b}), i.e., do not depend on $m$. In this case, we denote the hidden-symmetric measure as 
    \be
	\mathbb{P}^{(m)} = \mathbb{P}^{\infty},
	\label{eq_SHS1d}
    \ee
with the hidden scale invariance property
    \be
	\mathbb{P}^{\infty} = a^{\pm 1}_ \sharp \mathbb{P}^{\infty}.
	\label{eq_SHS1dB}
    \ee
%and satisfied the hidden scale invariance relations
 %   \be
%	\mathbb{P}^{\infty} = a^{\pm 1}_ \sharp \mathbb{P}^{\infty}.
%	\label{eq_SHS1d_HS}
 %   \ee
The infinity in the superscript reflects the inertial interval limit (\ref{eq:limits}). In fact, a proper mathematical formulation for the statistically restored hidden symmetry (\ref{eq_SHS1d}) would describe the convergence $\mathbb{P}^{(m)} \to \mathbb{P}^{\infty}$ in the limit (\ref{eq:limits}). 

The statistically restored hidden symmetry in the inertial interval is the central conjecture of our work, which we verify numerically and explore the implications thereof.
Being weaker than the  classical scaling symmetries (\ref{eqS1a}), the hidden  symmetry can be restored even when all classical scaling symmetries are broken. In this case, the hidden scale invariance has far reaching consequences, such as the universality of Kolmogorov multipliers, the anomalous scaling of structure functions, the fusion rules and others, as we demonstrate below.

\subsection{Universality of Kolmogorov multipliers}

\begin{figure}[htb]
\includegraphics[width=0.49\textwidth,trim=0cm 0cm 0cm 0cm, clip]{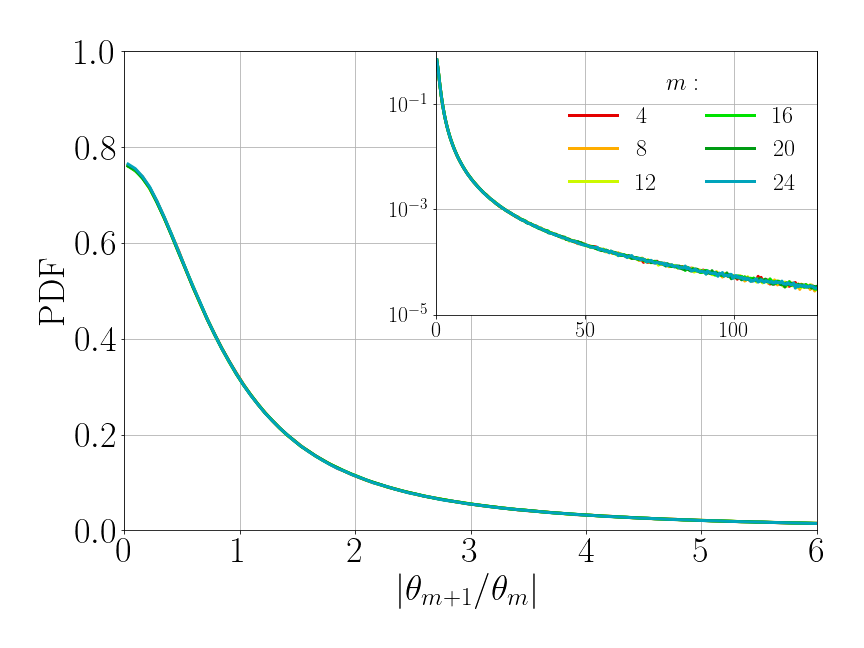}
\includegraphics[width=0.49\textwidth,trim=0cm 0cm 0cm 0cm, clip]{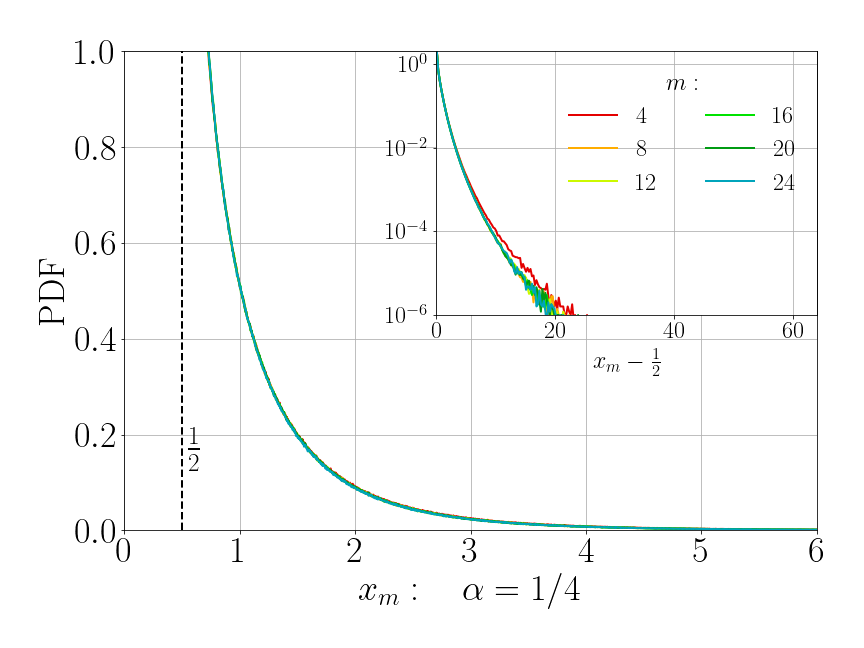}
\includegraphics[width=0.49\textwidth,trim=0cm 0cm 0cm 0cm, clip]{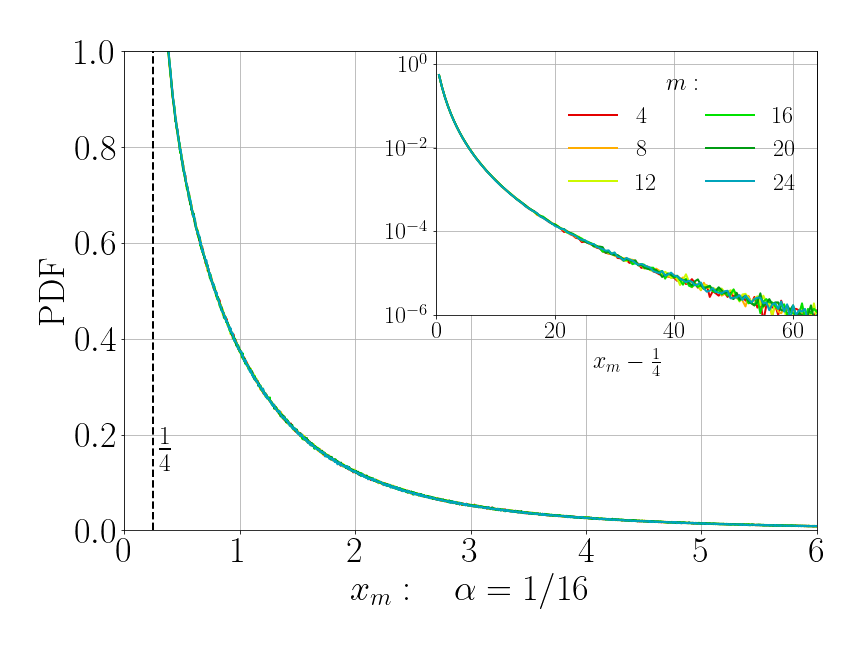}
\includegraphics[width=0.49\textwidth,trim=0cm 0cm 0cm 0cm, clip]{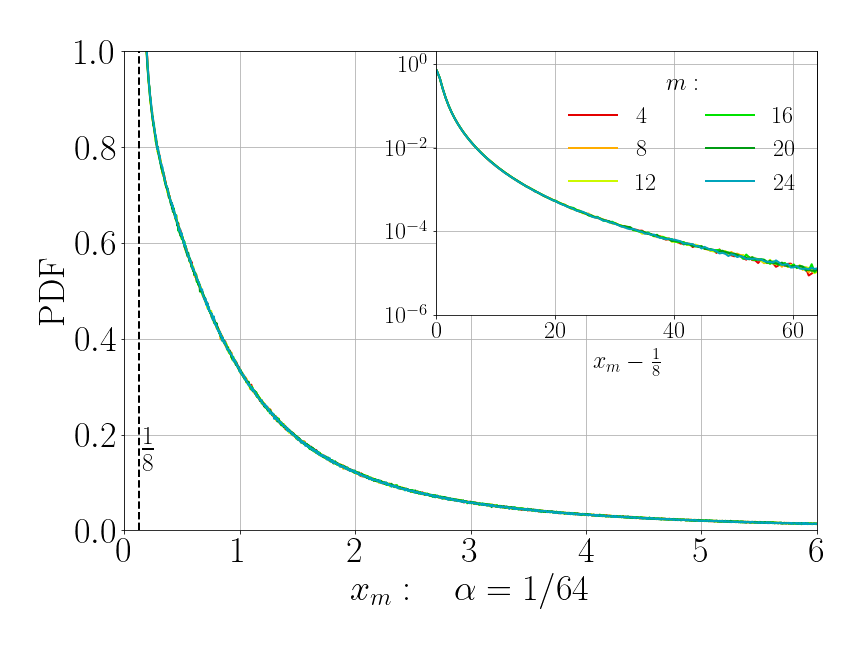}
\setlength{\unitlength}{\columnwidth}
\begin{picture}(1,0)(0,0)
\put(0.26,0.77){(a)}
\put(0.75,0.77){(b)}
\put(0.26,0.40){(c)}
\put(0.75,0.40){(d)}
\end{picture}
\caption{Collapse of PDFs for the multipliers (a) with the standard definition \eqref{eq:km} and  (b,c,d) with  the definition \eqref{eq:kmAx}
for various values of $\alpha$. Inset uses log scale for the $y$ axis.}
\label{fig:3}
\end{figure}
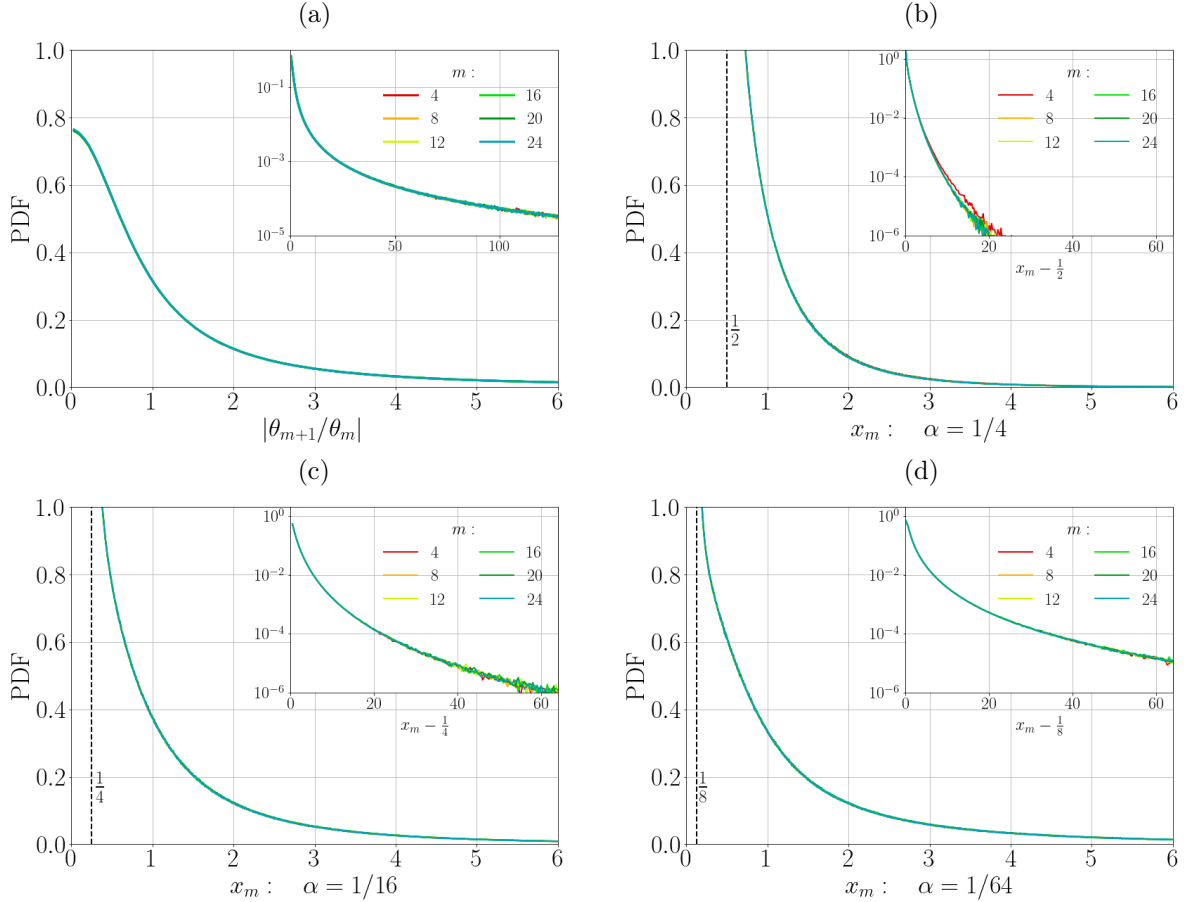

To evidence the statistical restoring of hidden symmetry, we follow  Kolmogorov's ideas of 1962~\cite{kolmogorov1962refinement} and consider the statistics of multipliers. In shell models, the multipliers are defined as the ratios $w_m = |\theta_m/\theta_{m-1}|$~\cite{benzi1993intermittency,eyink2003gibbsian}. Here, these can be expressed in terms of the rescaled variables \eqref{eqS3} as
    \be
	w_m(t) = \left|\dfrac{\theta_m(t)}{\theta_{m-1}(t)}\right| = \mathcal{W}_0[\Theta^{(m)}(\tau^{(m)})], \quad 
	\tau^{(m)} = \gamma^{2m} t, 
	\label{eq:km}
    \ee
with the function $\mathcal{W}_0[\Theta] = |\Theta_0/\Theta_{-1}|$.
Hence, the probability distributions of multipliers $w_m$ are given by the pushforward measures $\mathcal{W}_{0\sharp} \mathbb{P}^{(m)}$. The hypothesis of statistically restored hidden symmetry (\ref{eq_SHS1d}) implies that the distributions of multipliers are universal (independent of $m$) in the inertial interval. This feature is indeed verified in our numerical experiments.  Figure~\ref{fig:3}(a) shows  that the distributions nearly perfectly collapse  within the inertial interval. The  deviations start to emerge only at the cross-over with the forcing range.  This type of statistical universality of multipliers is similar to other shell models and the Navier--Stokes turbulence~\cite{mailybaev2021hidden,mailybaev2022hidden,mailybaev2022hiddenMT}, and strongly supports  the hypothesis of statistically restored hidden symmetry.  

More generally, the argument applies for any observable that can be expressed in terms of the rescaled variables.  
In particular, we later use  multipliers defined in terms of the scalar amplitudes as $x_m = \mathcal{A}_m[\theta]/\mathcal{A}_{m-1}[\theta]$. Using  Eqs.~(\ref{eqS2}) and (\ref{eqS3}) (see Appendix \ref{app:C}), these multipliers are expressed as 
  \be
  	x_m(t) = 
	\frac{\mathcal{A}_m[\theta(t)]}{\mathcal{A}_{m-1}[\theta(t)]}= \mathcal{X}_0[\Theta^{(m)}(\tau^{(m)})], 
	\quad 
	\tau^{(m)} = \gamma^{2m} t,
 	\label{eq:kmAx}
  \ee
with the function
  \be
	\mathcal{X}_0[\Theta] =\sqrt{\dfrac{\alpha}{1-{\Theta_0}^2}} \ge \sqrt \alpha.
 	\label{eq:kmAxF}
  \ee
The chief advantage of those new multipliers is that they are constructed as ratios between two strictly positive quantities. Their (single-shell) distributions  are given by the pushforward measures $\mathcal{X}_{0\sharp} \mathbb{P}^{(m)}$, whose universality (independence of $m$) is strongly supported by our numerics. Figures \ref{fig:3}(b--d)  show an accurate collapse of the PDFs, regardless of the specific value of $\alpha$ entering the definition of the scalar amplitudes \eqref{eqS2}. Similar arguments apply to the universality of joint distributions for multipliers.

%##############################################
%PERRON-FROBENIUS
\section{Perron--Frobenius scenario to anomalous scaling laws}
\label{seq:Perron}
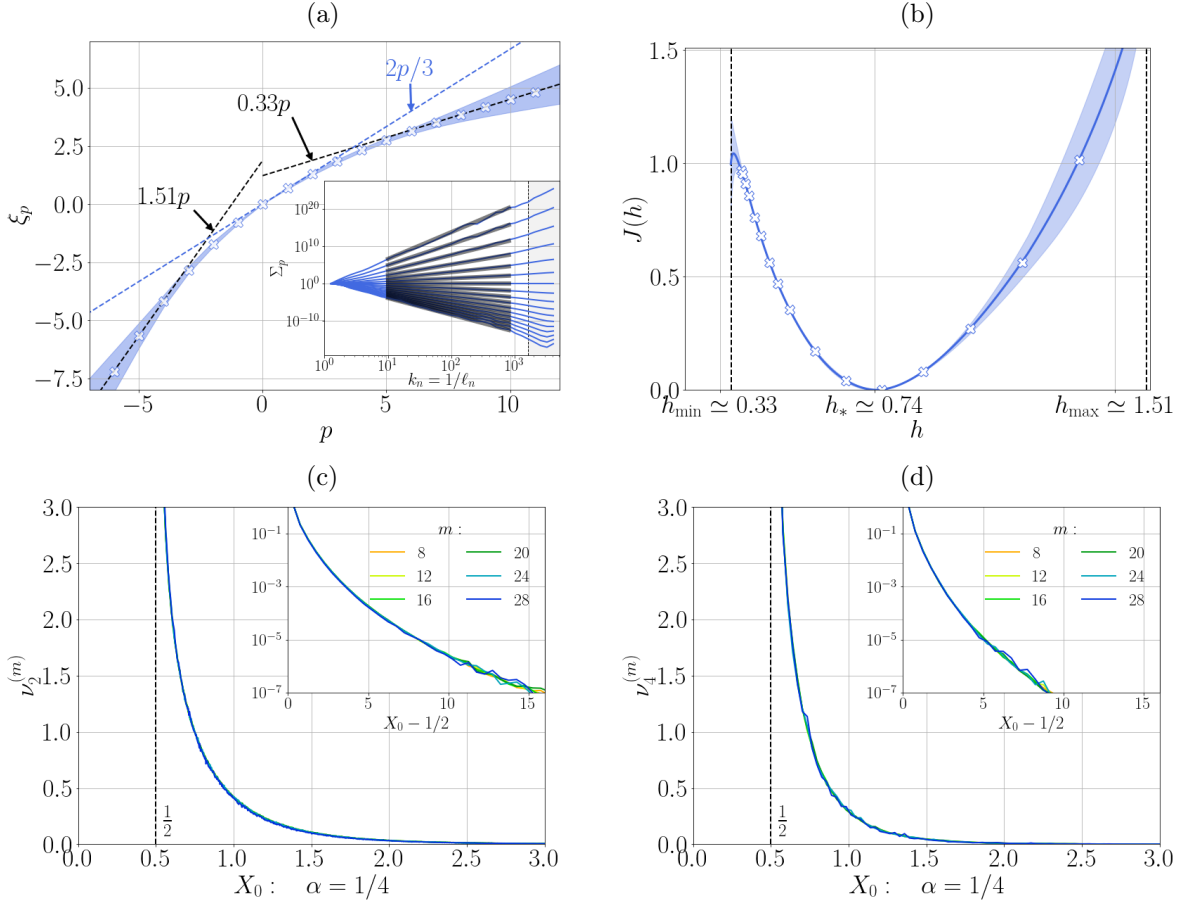
\begin{figure}[bth]
\includegraphics[width=0.49\textwidth,trim=0cm 0cm 0cm 0cm, clip]{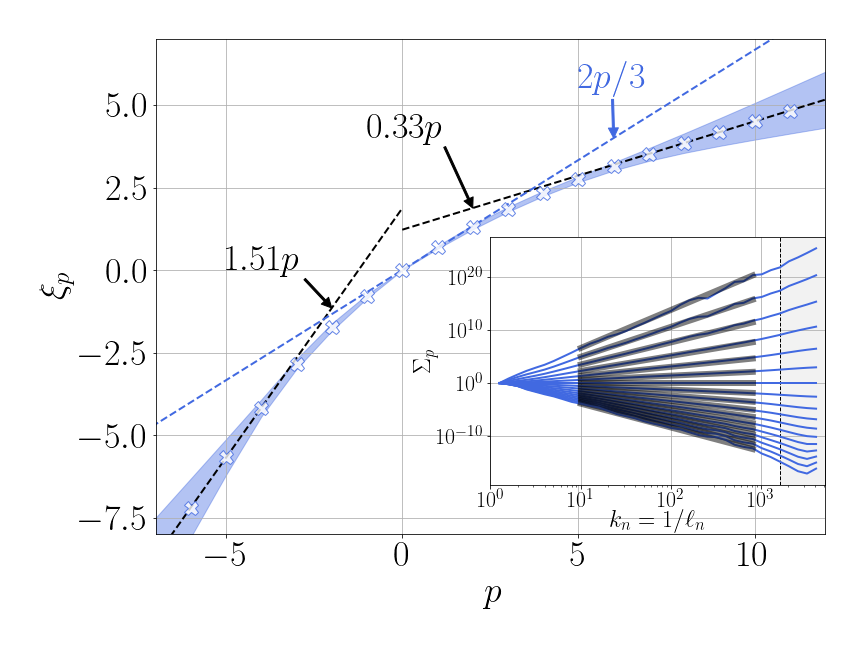}
\includegraphics[width=0.49\textwidth,trim=0cm 0cm 0cm 0cm, clip]{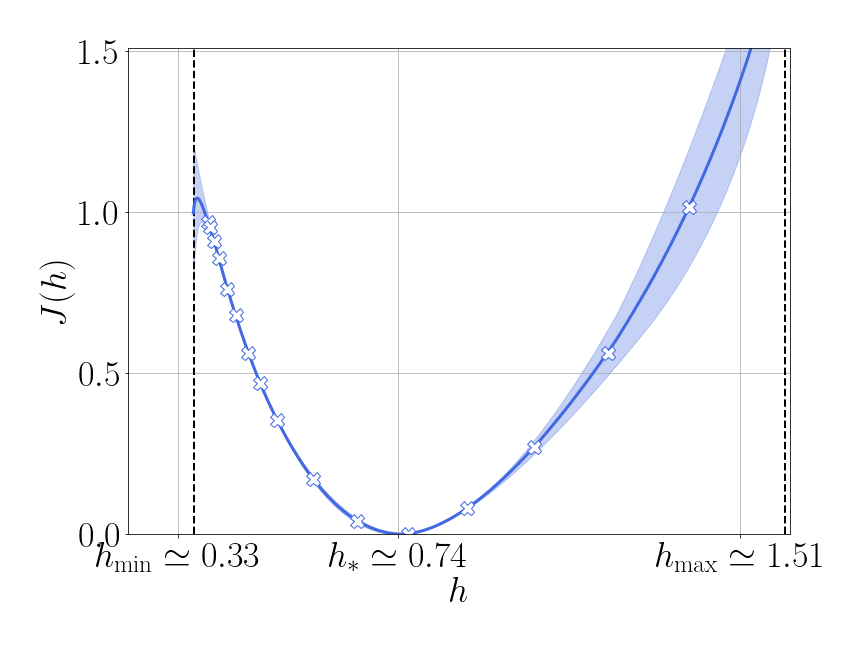}
\includegraphics[width=0.49\textwidth,trim=0cm 0cm 0cm 0cm, clip]{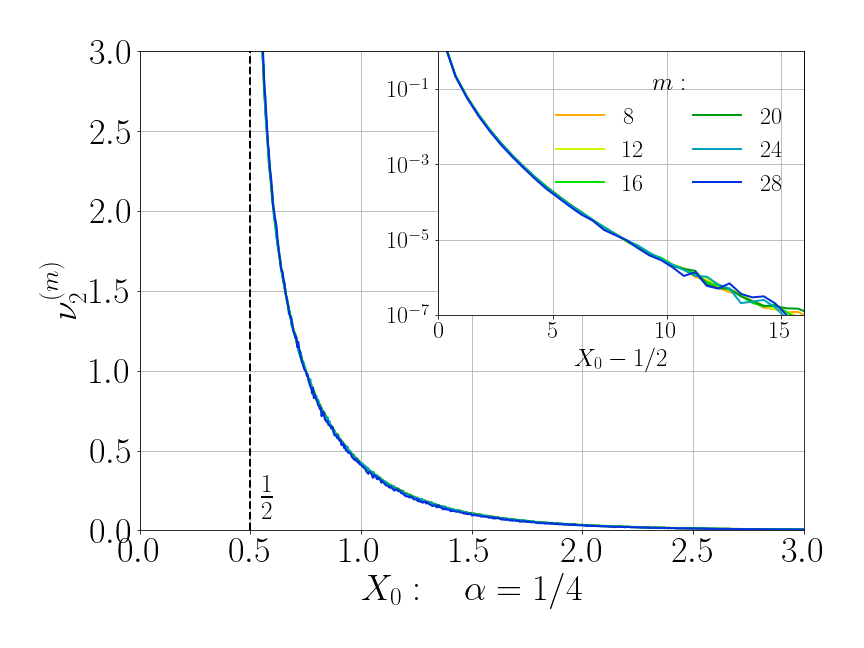}
\includegraphics[width=0.49\textwidth,trim=0cm 0cm 0cm 0cm, clip]{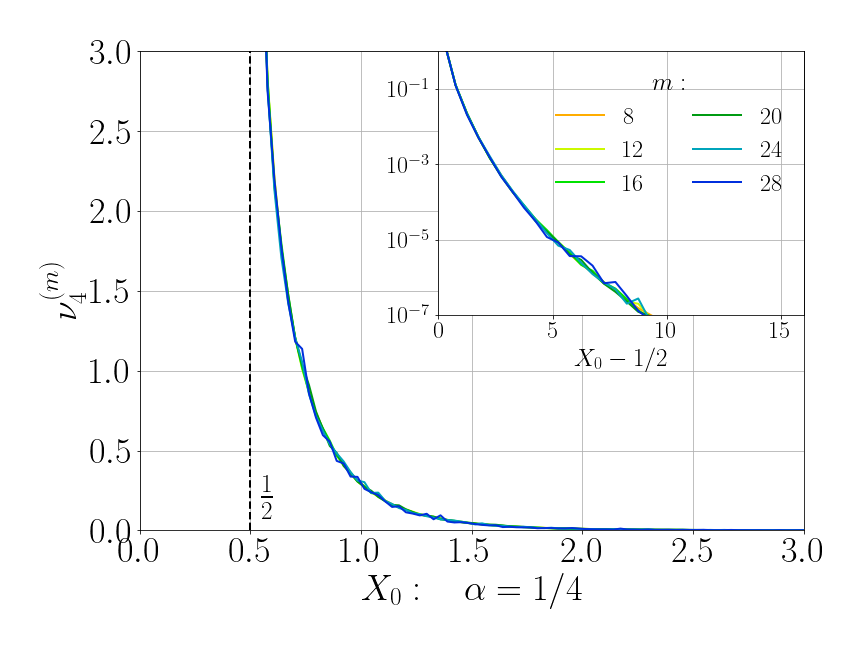}
\setlength{\unitlength}{\columnwidth}
\begin{picture}(1,0)(0,0)
\put(0.27,0.77){(a)}
\put(0.75,0.77){(b)}
\put(0.27,0.395){(c)}
\put(0.75,0.395){(d)}
\end{picture}
\vspace{-2cm}
\caption{(a) The scaling exponents $\xi_p$ for the structure function~\eqref{eq:SpA}
based on scalar amplitudes with $\alpha =1/4$%, indicated with white dots. The white crosses indicate the exponents $\zeta_p$ for the usual structure functions from Fig. \ref{fig:1}
. 
Insets show the scaling laws for even values of $p$ from $p = -6$ (top) to $p=10$ (bottom).  (b) The corresponding rate function  $J(h)$.
(c,d) Collapse of marginal measure  $\nu^{(m)}_p$ from Eq.~\eqref{eq:fp_v1} for respectively $p= 2$ and $p=4$.}
\label{fig:4}
\end{figure}

The zero-mode analysis of \S \ref{ssub:zeromode}  highlighted the degeneracy of the scaling symmetry \eqref{eqS1a}, through
 anomalous power law scalings for the structure functions 
$\mS_p$ of orders  $p=2$ and $4$. We now intend to address the general case with $p \in \mathbb R$.    As the averages \eqref{eq:Sp} diverge for large negative orders because of the shell variables accidentally vanishing, we here prefer to work with the moments of scalar amplitudes defined as
   \begin{equation}
    \Sigma_p (\ell_n) = \lim_{T\to \infty}T^{-1}\int_0^ T \av{ \mathcal{A}_m^p[\theta] } dt
    \quad  \text{for} \quad p \in \mathbb{R}.
    \label{eq:SpA}
    \end{equation}
Unlike Eq.~\eqref{eq:Sp}, we will argue that for suitable values of the parameter $\alpha$ in expression (\ref{eqS2}), the functions (\ref{eq:SpA}) allow to consider negative moments. As such, they prove to be more convenient for both mathematical and numerical analysis.
Besides, as shown in Fig. \ref{fig:4}(a), the  structure functions measured in our numerical experiments demonstrate (anomalous) power-law scalings $\Sigma_p(\ell_n) \propto \ell_n^{\xi_p}$, in the inertial interval for both positive and negative orders $p$. For positive orders, we have 
$\xi_p=\zeta_p$, coinciding  with the scaling exponents  of the usual moments \eqref{eq:Sp} shown in Fig. \ref{fig:1}(b). This identification will be discussed in \S\ref{sec:6}.
In this section, we show that the power-law scalings with anomalous exponents follow from the statistical hidden symmetry, emerging as a suitably defined Perron-Frobenius mode. 

%%%%%%%%%%%%%%%%%%%%%%%%%%%%%%%%%%%%%%%%%%%%%%%%%%%%%%%
\subsection{Structure functions as iterated measures on multipliers}
In this subsection, we give a technical derivation showing that the computation of structure functions reduces to an iteration procedure, mediated by linear operators determined by the hidden-symmetry transformation. %Later, we relate the scaling exponents $\xi_p$ to eigenvalues of these operators.
To that end, we observe that the scalar amplitude is nothing but the telescopic product of multipliers (\ref{eq:kmAx}):
\be
 \mathcal{A}_m[\theta(t)] = A_0\prod_{n = 1}^{m} x_n(t),
\label{eq:telescopprod}
\ee 
with the constant prefactor $A_0 = \mathcal{A}_0[\theta] = 1/\sqrt{1-\alpha}$. At this point, please observe that the product
in Eq.~(\ref{eq:telescopprod}) extends way beyond the inertial range of scales, as it involves multipliers within the forcing range at scales $\ell_n \sim \ell_0$. Hence, one should be careful with  the hypothesis of statistically restored hidden symmetry (\ref{eq_SHS1d}): Being  restricted to the inertial interval, it does not apply at the level of the scalar amplitude \eqref{eq:telescopprod}.
%It is 
%We now express the multipliers in Eq.~(\ref{eq:telescopprod}) and their time averages in terms of the rescaled field $\Theta^{(m)}$ and its probability measure $\mathbb{P}^{(m)}$. 
%Let us here recall that whole theory of Section~\ref{sec3} was developed assuming that the variables belong to the inertial interval. This is not the case in Eq.~(\ref{eq:telescopprod}), which contains multipliers from the forcing range at large scales $\ell_n \sim \ell_0$. One can check that the hidden symmetry transformations (\ref{eq:HS}), which map the rescaled processes $\Theta^{(m)}(\tau^{(m)})$ into $\Theta^{(m\pm1)}(\tau^{(m\pm1)})$, are the consequences of the definition (\ref{eqS3}) alone (see Appendix~\ref{app:C}) and, therefore, remain valid in the forcing range. One should be careful, however, with the concept of the statistically restored hidden symmetry (\ref{eq_SHS1d}), which is restricted to the inertial interval.
%
The specific forms  (\ref{eq:HS}) of the hidden symmetry shift, which map the rescaled processes $\Theta^{(m)}(\tau^{(m)})$ into $\Theta^{(m\pm1)}(\tau^{(m\pm1)})$, are more general. They are the consequences of the definition (\ref{eqS3}) alone  and, therefore, remain valid in the forcing range -- see Appendix~\ref{app:C} for more details.

Iterating the hidden symmetry transform  (\ref{eq:HS}) yields $\Theta^{(m+N)} = a^{N} [\Theta^{(m)}]$. Combined with  Eq.~(\ref{eq:kmAx}), this prescribes the multipliers $x=(x_n)_{n\in\mathbb Z}$ in terms of the maps as
  \be
  	x_n = \mathcal{X}_0[\Theta^{(m+N)}] 
	=  \mathcal{X}_N [\Theta^{(m)}], 
	\quad
	\mathcal{X}_N = \mathcal{X}_0 \circ a^N, 
	\quad n = m+N.
 	\label{eq:kmAx_J}
  \ee
For simplicity, we dropped the corresponding time arguments as they do not affect time averages and, hence, the statistical properties. 
 The pushforward by the map $\mX[\Theta] = (\mX_N[\Theta])_{N \in \mathbb{Z}}$ defines the probability measure $\mathbb{P}_{\mX}^{(m)} = \mX_{\sharp} \mathbb{P}^{(m)}$ in the space of infinite-dimensional sequences $X = (X_N)_{N \in \mathbb{Z}}$ with positive components $X_N > 0$.

We now use the notation $X_\ominus = (X_0,X_{-1},X_{-2},\ldots)$ for inverse sequences of real positive numbers. As such,
$\mathbb{P}_{\mX}^{(m+1)}(dX_\ominus) $ denotes the measure $\mathbb{P}_{\mX}^{(m+1)}$ restricted to the subspace $X_{\ominus}$, which in practice prescribes the joint multiplier statistics  $(x_m,x_{m-1},x_{m-2}, \ldots)$.
% and the notation $\mathcal{X}_\ominus[\Theta] = (\mathcal{X}_0[\Theta],\mathcal{X}_{-1}[\Theta],\ldots)$ for the inverse sequence of functions (\ref{eq:kmAx_J}). 
%The pushforward by the map $X_\ominus = \mathcal{X}_\ominus[\Theta]$ defines the probability measure $\mathbb{P}_{X_\ominus}^{(m)} = \mathcal{X}_{\ominus \sharp} \mathbb{P}^{(m)}$ in the space $X_\ominus$. 
From Eq.~(\ref{eq:kmAx_J}), we observe that the change $m \mapsto m+1$ reduces to a simple shift $N \mapsto N-1$. As a consequence, using Bayes' formula defining conditional probabilities \cite{evans2012introduction}, the measures $\mathbb{P}_{\mX}^{(m)}$ and $\mathbb{P}_{\mX}^{(m-1)}$ obey the recursion relation
   	\begin{equation}
	\mathbb{P}_{\mX}^{(m)}(dX_\ominus) = 
	\mathbb{P}_{\mX}^{(m)}(dX_0| X_-)\, 
	\mathbb{P}_{\mX}^{(m-1)}(d\tilde{X}_\ominus),
    	\label{eq:SpAm_P}
    	\end{equation}
where $\tilde{X}_\ominus = (\tilde{X}_0,\tilde{X}_{-1},\ldots)$ with $\tilde{X}_N = X_{N-1}$, and $\mathbb{P}_{\mX}^{(m)}(dX_0| X_-)$ denotes the conditional measure for the first component of the sequence $X_\ominus = (X_0,X_-)$ with $X_- = (X_{-1},X_{-2},\ldots)$. 

Using the ergodicity assumption in Eq.~(\ref{eq:SpA}) with the expressions (\ref{eq:telescopprod}) and (\ref{eq:kmAx_J}), we have 
   	\begin{equation}
    	\Sigma_p(\ell_m) = A_0^p \, \mathbb{E}_{\mX}^{(m)}\left( \prod_{N = 1-m}^0  X_N^p \right)=\int \mu_p^{(m)}(dX_\ominus),
    	\label{eq:SpAm}
    	\end{equation}
involving the  positive  measures $\mu_p^{(m)}(dX_\ominus)$ defined in the space $X_\ominus$
as
	\be
%    	\Sigma_p(\ell_m) = \int \mu_p^{(m)}(dX_\ominus),\quad
	\mu_p^{(m)}(dX_\ominus) = A_0^p \, \left( \prod_{N = 1-m}^0  X_N^p \right) 
	\mathbb{P}_{\mX}^{(m)} (dX_\ominus).
	\label{eq:mump}
	\ee
We point out that the mesures $\mu_p^{(m)}$ are   not probability measures as they do not generally have unit mass.
Their salient feature is that they obey a recursion relation in $m$, which we now derive. 

%where we introduced a positive (but not probability) measure $\mu_p^{(m)}(dX_\ominus)$ in the space $X_\ominus$. 
%Using Eqs.~(\ref{eq:SpAm_P}) and (\ref{eq:mump}) we write the relation

Using Eq.~(\ref{eq:mump}) and expressing the measure from Eq.~(\ref{eq:SpAm_P}) with $\tilde{X}_N = X_{N-1}$ yields
   	\begin{equation}
	\begin{array}{rcl}
	\mu_p^{(m)}(dX_\ominus) 
	& = & \displaystyle
	X_0^p \,\mathbb{P}_{\mX}^{(m)}(dX_0| X_-) \left[ A_0^p \, \left( \prod_{N = 2-m}^0  \tilde{X}_N^p \right) 
	\mathbb{P}_{\mX}^{(m-1)} (d\tilde{X}_\ominus) \right] \\[10pt]
	& = & \displaystyle
	X_0^p \,
	\mathbb{P}_{\mX}^{(m)}(dX_0| X_-)\, 
	\mu_p^{(m-1)}(d\tilde{X}_\ominus).
	\end{array}
    	\label{eq:SpAm_PIt}
    	\end{equation}
%We will use an equivalent representation of Eq.~(\ref{eq:SpAm_PIt}) as
 %  	\begin{equation}
%	\mu_p^{(m+1)} = \mathcal{L}_p^{(m+1)} [\mu_p^{(m)}],
 %   	\label{eq:SpAm_PL}
  %  	\end{equation}
In a more compact but equivalent form we write
   	\begin{equation}
	\mu_p^{(m)}=\mathcal{L}_p^{(m)} [\mu_p^{(m-1)}], \quad \mathcal{L}_p^{(m)} : \mu(dX_\ominus) \mapsto
	X_0^p \,
	\mathbb{P}_{\mX}^{(m)}(dX_0| X_-)\, 
	\mu(d\tilde{X}_\ominus),
    	\label{eq:SpAm_Linf}
    	\end{equation}
involving $\mathcal{L}_p^{(m)}$ as  linear positive operator acting on positive measures $\mu(dX_\ominus)$.
The iterative use of Eq.~(\ref{eq:SpAm_Linf}) yields the explicit composition representation
   \begin{equation}
	\mu_p^{(m)} = \mathcal{L}^{(m)}_p \circ \mathcal{L}^{(m-1)}_p \circ \cdots \circ 
	\mathcal{L}^{(1)}_p [\mu_p^{(0)}],
    \label{eq_Sp_yE2}
    \end{equation}
where $\mu_p^{(0)}(dX_{\ominus}) = A_0^p \, \mathbb{P}_{\mX}^{(0)}(dX_{\ominus})$ is just the Dirac measure of constant multipliers $x_n = X_N = 1$ for $n = N \le 0$, as prescribed by the (large-scale) boundary conditions in Eq.~\eqref{eq4}.

%%%%%%%%%%%%%%%%%%%%%%%%%%%%%%%%%%%%%%%%%%%%%%%%%%%%%%%
\subsection{Scaling exponents as  Perron--Frobenius eigenvalues}
As we already mentioned, the statistical recovery of hidden symmetry (\ref{eq_SHS1d}) cannot be applied directly to the measures $\mu_p^{(m)}$ from Eq.~(\ref{eq:mump}), because these measures depend on the large-scale statistics of the forcing range. Instead, we use the statistical recovery of hidden symmetry for the linear operators (\ref{eq:SpAm_Linf}), which depend only on the conditional measure $\mathbb{P}_{\mX}^{(m)}(dX_0| X_-)$. Indeed, this conditional measure is restricted to the inertial interval provided that the shell $m$ belongs to the inertial interval and the correlations of multipliers are local. Therefore, within the inertial interval we have the asymptotic relation 
   \begin{equation}
   \mathcal{L}^{(m)}_p = \mathcal{L}^{\infty}_p, \quad
   	\mathcal{L}_p^{\infty} : \mu(dX_\ominus) \mapsto
	X_0^p \,
	\mathbb{P}_{\mX}^{\infty}(dX_0| X_-)\, 
	\mu(d\tilde{X}_\ominus),
    \label{eq_Sp_yE4}
    \end{equation}
with the universal (independent of $m$) linear operator $\mathcal{L}^{\infty}_p$.

The linear operator $\mathcal{L}_p^{\infty}$ is positive in the sense that it maps positive measures to positive measures. It follows that its spectral radius is given by a real positive (Perron--Frobenius) eigenvalue $\lambda_p$ satisfying the eigenvalue problem~\cite{lax2007linear,deimling2010nonlinear}
	\begin{equation}
	\mathcal{L}_p^{\infty} [\nu_p] = \lambda_p \,\nu_p.
	\label{eq:Perron}
	\end{equation}
Here, the eigenvector $\nu_p(dX_\ominus)$ is a positive measure which we normalize to have unit mass $\int \nu_p = 1$.
Under technical non-degeneracy assumptions (referring to strict positivity and compactness~\cite[Sec. 19.5]{deimling2010nonlinear}), the Perron-Frobenius eigenvalue $\lambda_p$ is larger than the absolute values of all remaining eigenvalues.
 Combining Eq.~(\ref{eq_Sp_yE2}) with (\ref{eq_Sp_yE4}) and (\ref{eq:Perron}), we conclude that the measures $\mu_p^{(m)}$ from the inertial interval (for large shells $m$) have the asymptotic form 
	\begin{equation}
	    	\mu_p^{(m)}  = c_p \lambda_p^m \, \nu_p,
    	\label{eq:PerronAsymp}
    	\end{equation}
involving  a positive coefficient $c_p$ independent of $m$.
Substituting expression \eqref{eq:PerronAsymp} into \eqref{eq:SpAm}, we obtain  asymptotic power laws for the structure functions as
	\begin{equation}
	\Sigma_p(\ell_m) = c_p \lambda_p^m 
	= c_p \left(\frac{\ell_m}{\ell_0}\right)^{\xi_p},\quad 	\xi_p = -\log_\lambda \lambda_p,
    	\label{eq1_SpH11}
    	\end{equation}
where we expressed $\ell_m/\ell_0 = \lambda^{-m}$.
The scaling exponent is now obtained explicitly in terms of the Perron--Frobenius eigenvalue.
This relation allows (and yields in general) a nonlinear dependence of the exponents $\xi_p$ on the order $p$~\cite{mailybaev2022hidden}. In this case, the hidden scale invariance provides anomalous power-law scalings for structure functions, and therefore prescribes the intermittency. 
We remark that, according to Eq.~(\ref{eq_Sp_yE2}), the pre-factors $c_p$ are determined by the large-scale statistics, i.e., by the forcing conditions. 

%%%%%%%%%%%%%%%%%%%%%%%%%%%%%%%%%%%%%%%%%%%%%%%%%%%%%%%
\subsection{Numerical verification of the Perron--Frobenius scenario} 

The convergence (up to a constant prefactor) of the measures $\mu_p^{(m)}$ towards an eigenvector measure $\nu_p$ in Eq.~(\ref{eq:PerronAsymp}) can be addressed numerically. This provides a further assessment of the statistical hidden scale invariance and the confirmation of the Perron--Frobenius scenario for the anomalous scaling. 
To that end, we normalize the measure $ \mu_p^{(m)}(dX_\ominus)$ to unit mass and introduce the corresponding marginal distribution as
	\begin{equation}
	\nu_p^{(m)}(dX_0) 
	= \frac{1}{\Sigma_p(\ell_m)} \int_{X_-} \mu_p^{(m)}(dX_0,dX_-).
    	\label{eq:fp_v1}
    	\end{equation}
The Perron-Frobenius scenario predicts that the measures $\nu_p^{(m)}(dX_0)$ should converge towards the single-shell marginal of the eigenvector $ \nu_p (dX_0 )$, which by construction has unit mass and is independent of $m$.

From a numerical standpoint,  the distribution \eqref{eq:fp_v1} is estimated as the limit
	\begin{equation}
	\nu_p^{(m)}(dX_0) 
	= \dfrac{1}{\Sigma_p(\ell_m)} \lim_{T\to \infty}T^{-1}\int_0^ T dt\, \left\langle \mathcal{A}_m^p[\theta] 
	\mathbbm{1}_{x_m(t) \in dX_0}\right\rangle,
    	\label{eq:fp_v1B}
    	\end{equation}
which stems from the definition (\ref{eq:mump}) combined with the expressions (\ref{eq:SpA}) and (\ref{eq:kmAx_J}).
Outcomes from our numerics  are presented in Fig.~\ref{fig:4}(c,d) for $p = 2, 4$. Each panel shows the densities for the shells $m = 4,\ldots,28$ in the inertial interval. The accurate collapse of the graphs verifies the
convergence of the measures $\nu_p^{(m)}$  towards a well-identified (Perron-Frobenius) eigenmode and
provides strong numerical support to our theory.
%%%%%%%%%%%%%%%%%%%%%%%%%%%%%%%%%%%%%%%%%%%%%%%%%%%%%%%%%%%
% FROM HIDDEN SYMMETRY TO ZERO MODES
\corr 
\section{From Perron-Frobenius to fusion rules and multifractality}
\rroc
\label{sec:6}
This section discusses the implications of  the Perron-Frobenius scenario  beyond the observation of power-law scaling for the   specific structure functions  \eqref{eq:SpA}, based on the scalar amplitudes. We show that
the statistical restoring of hidden symmetry also prescribes: 
\emph{(i)} the scaling of usual structure functions $\mS_p$ with the same exponents, at least for the orders when the time average is well-defined. \emph{(ii)} the fusion rules required to 
evidence the zero-mode scenario and \emph{(iii)}  the Parisi--Frisch multifractal formalism.

%%%%%%%%%%%%%%%%%%%%%%%%%%%%%%%%%%%%%
\subsection{Scaling of the usual structure functions}
Through  Eq.~(\ref{eq:PerronAsymp}),  the Perron-Frobenius scenario predicts  power-law scalings for the measures $\mu_p^{(m)}$, and hence for any observables that can be associated with these measures. Scalings for the structure functions $\Sigma_p$ in Eq.~(\ref{eq1_SpH11}) represent just a particular example. We now show that the scalings for usual structure functions $\mS_p$ from Eq.~(\ref{eq:Sp}) follow in a similar manner.

Using Eqs. \eqref{eqS2}, \eqref{eq:kmAx} and (\ref{eq:telescopprod}), one obtains the identity
	\begin{equation}
 	\label{eqZM1}
	|\theta_m| 
	= \sqrt{\mathcal{A}_m^2[\theta]-\alpha \mathcal{A}_{m-1}^2[\theta]}
	= \sqrt{1-\frac{\alpha \mathcal{A}_{m-1}^2[\theta]}{\mathcal{A}_m^2}}\, \mathcal{A}_m
	= \sqrt{1-\frac{\alpha}{x_m^2}}\, A_0 \prod_{n = 1}^{m}x_n.
	\end{equation}
This expression differs from Eq.~(\ref{eq:telescopprod}) only by the square root prefactor depending on $x_m$. One can check that this prefactor 
modifies the expression (\ref{eq:SpAm}) for the structure function (\ref{eq:Sp}) as
	\be
	\mS_p(\ell_m) = \int 
	\left( 1-\frac{\alpha}{X_0^2} \right)^{p/2} \mu_p^{(m)}(dX_\ominus).
 	\label{eqZM3x}
    	\ee
Finally, using the Perron--Frobenius asymptotic relation (\ref{eq:PerronAsymp}), we obtain 
\be
	\mS_p(\ell_m) = \tilde{c}_p \left(\frac{\ell_m}{\ell_0}\right)^{\xi_p},\quad 
	\tilde{c}_p =  c_p \int  \left(1-\frac{\alpha}{X_0^2}\right)^{p/2}\, \nu_p(dX_\ominus),
    	\label{eq1_SpH11xx}
\ee
involving  the same exponent $\xi_p = -\log_\lambda \lambda_p$ but a different prefactor $\tilde c_p$.
 Naturally, the relation \eqref{eq1_SpH11xx} is meaningful only provided that the integral in the second expression  converges. For the structure functions $\mS_p$, the divergence is expected for negative orders $p \le -1$ due to accidental zeros of shell variables. 
Our numerics supports this scenario: for $p>-1$, the exponents $\zeta_p=\xi_p$, meaning that the scaling exponents of  $\mS_p$ and $\Sigma_p$ coincide. Figure~\ref{fig:1}(b) and \ref{fig:4}(a) show that both families share the same  asymptotic behavior 
\be
	\xi_p=\zeta_p \sim 0.33 p \quad \textrm{as} \quad p \to +\infty.
     	\label{eq1_Sp_pInf}
\ee
Divergence are seen for $ p\le -1$:   The usual structure functions $\mS_p$  diverge while their scalar amplitude counterparts of  Fig.~\ref{fig:4}(a) are well-behaved, with the apparent asymptotic 
\be
	\xi_p \sim 1.51 p \quad \textrm{as} \quad p \to -\infty.
     	\label{eq1_Sp_pMInf}
\ee
\corr We remark, however, that the linear behaviors (\ref{eq1_Sp_pInf}) and (\ref{eq1_Sp_pMInf}) at large orders  appear in the regions of increasing statistical errors, which may question their validity. Similar properties are observed in other shell models; see e.g. \cite{l2001outliers,mailybaev2023hidden,de2024extreme}.\rroc

%%%%%%%%%%%%%%%%%%%%%%%%%%%%%%%%%%%%%%%%%%%%%%%%
\subsection{Fusion rules}
The hidden scaling symmetry determines the fusion rules formulated for the fourth-order correlators by Eq.~(\ref{eqZ6}). It is sufficient to prove that
	\begin{equation}
 	\label{eqZ6n}
	\sigma_{m(-l)} = C_{-l} \sigma_{m0} \quad \textrm{for} \quad l \ge  0.
	\end{equation}
Indeed, relation (\ref{eqZ6n}) yields Eq.~(\ref{eqZ6}) with a different sign of $l$ as
	\begin{equation}
	\sigma_{nl} = 
	\sigma_{(n+l)(-l)} =  
	C_{-l} \sigma_{(n+l)0} = 
	C_{-l} \lambda^{-\zeta_4l} \sigma_{n0} = 
	C_{l} \sigma_{n0}, \quad 
	C_l = C_{-l}\lambda^{-\zeta_4l},
 	\label{eqZ6npr}
	\end{equation}
where we also used Eqs.~(\ref{eq:4order}) and (\ref{eq7B3}).
Using Eq.~(\ref{eqZM1}) for $m$ and $m-l$, we express
	\begin{equation}
 	\label{eqZM2}
	\theta_m^2\theta_{m-l}^2 
	= \left(1-\frac{\alpha}{x_m^2}\right)\left(1-\frac{\alpha}{x_{m-l}^2}\right)
	\left(\prod_{j = m-l+1}^{m}x_j^{-2} \right)
	\mathcal{A}_m^4[\theta].
	\end{equation}
%In addition to $\mathcal{A}_m^4[\theta]$ from Eq.~(\ref{eq:telescopprod}), expression (\ref{eqZM2}) contains the prefactor depending on $x_{m-l+1},\ldots,x_m$. 
As in the previous subsection, one can check that the average of function (\ref{eqZM2}) yields the expression analogous to Eq.~(\ref{eq:SpAm}) but with the modified prefactor as
	\be
	\sigma_{m(-l)} = \int 
	\left(1-\frac{\alpha}{X_0^2}\right)\left(1-\frac{\alpha}{X_{-l}^2}\right)
	\left(\prod_{j = 0}^{l-1}X_{-j}^{-2} \right)
	\mu_4^{(m)}(dX_\ominus).
 	\label{eqZM3xy}
    	\ee
Using the Perron-Frobenius representation~(\ref{eq:PerronAsymp}), we find 
\be
	\sigma_{m(-l)} = \hat{c}_{l} \left(\frac{\ell_m}{\ell_0}\right)^{\xi_4},\quad 
	\hat{c}_{l} = c_4 \int 
	\left(1-\frac{\alpha}{X_0^2}\right)\left(1-\frac{\alpha}{X_{-l}^2}\right)
	\left(\prod_{j = 0}^{l-1}X_{-j}^{-2} \right)
	\nu_4(dX_\ominus).
    	\label{eq1_SpFR}
\ee
\corr This expression yields the fusion rule (\ref{eqZ6n}) with the coefficients $C_{-l} = \hat{c}_{l}/\hat{c}_0$.\rroc
Thus, the practical implementation of  the zero-mode theory follows from the Perron--Frobenius scenario and the statistical recovery of hidden scale invariance.

\subsection{Multifractal formalism}
\label{sec:multifractal}
The multiplicative representation of the scalar amplitudes  \eqref{eq:telescopprod} and the Perron-Frobenius scenario for intermittency provide an explicit multifractal interpretation of the scaling exponents through  large deviations, in particular the
 G\"artner--Ellis Theorem \cite{touchette2009large}. 
For positive $m$, let us define the effective H\"older exponents as the sample mean
\be
	H_m = - \dfrac1 {m}  \sum_{J=0}^{m-1} \log_\lambda X_{-J}, 
	%\quad \text{ with } X\overset{\text{Law}}\sim \mathbb P_\mX^{(m)},
	\label{eq:Hm}
\ee
where $X_0,X_{-1},\ldots$ are random numbers with the probability distribution $\mathbb P_\mX^{(m)}$. It follows from Eqs.~(\ref{eq:telescopprod}) and (\ref{eq:kmAx_J}) that $H_m = h$ identifies to the monofractal asymptotics $\mathcal{A}_m = A_0\lambda^{-mh}$. Besides, the characteristic function of $H_m$ recovers the structure functions through the following calculation
\be
	  \mathbb E_\mX^{(m)} \left[A_0^p\lambda^{-m p H_m}\right] 
	  = \Sigma_p(\ell_m) \sim c_p \lambda^{-m\xi_p}, \quad m\to \infty,
\ee
with the second equality stemming from the  Perron-Frobenius identity  \eqref{eq1_SpH11}. In other words, the Perron-Frobenius scenario provides existence of the  generating function 
\be
	p \mapsto -\lim_{m\to \infty}\dfrac{1}{m}\log_\lambda  \mathbb E_\mX^{(m)}\left[\lambda^{-m p H_m}\right]= \xi_p, \quad p \in \mathbb R.
\ee
Further assuming that the generating function is smooth for all real orders $p$, 
%, as suggested by the numerics presented in Fig.~\ref{fig:4}(a), 
the G\"artner--Ellis Theorem \cite{touchette2009large,mailybaev2023hidden} yields the large deviation estimate valid for  $m\to \infty$ as
\be
	\mathbb P_\mX^{(m)} \left(H_m \in dh\right) \approx \lambda^{-m J(h)} dh
	\label{eq:MF}
\ee
with the rate function
\be
	J(h)=\sup_{p \in \mathbb R} \, (\xi_p-ph ).
	\label{eq:MFR}
\ee
In the multifractal language \cite{frisch1980fully,frisch1995turbulence}, $J(h)$ plays the role of fractal codimension. So, the fractal dimension (singularity spectrum) is defined as $D(h) = d-J(h)$, where $d$ is the dimension of physical space. In turn, the scaling exponents are expressed through the Legendre--Fenschel transform as
    \begin{equation}
	    \xi_p = \inf_{h \in \mathbb{R}} \, \left[ph+J(h)\right].
	\label{eq_MF4}
    \end{equation}
Equations~\eqref{eq:Hm}, \eqref{eq:MF} and \eqref{eq_MF4} summarize the multifractal theory as a large deviation principle stemming from  the statistically recovery of the hidden symmetry.

From a numerical viewpoint, we infer from Fig.~\ref{fig:4}(a) that the function $\xi_p$  is apparently smooth and concave on $\mathbb R$ with linear asymptotics (\ref{eq1_Sp_pInf}) and (\ref{eq1_Sp_pMInf}).
It follows from classical properties of the Legendre--Fenschel transforms \cite{touchette2005legendre},
that the rate function is convex and compactly supported on
    \begin{equation}
	h_{\min} \le h \le h_{\max}, \quad 
	h_{\min} \approx 0.3, \quad 
	h_{\max} \approx 1.5,
    \label{eq_Lminmax}
    \end{equation}
with vertical asymptotics at $h_{\min}$ and $ h_{\max}$ -- see Fig. \ref{fig:4}(b).
As a final comment, please observe that the exponent $h_{\max}$ prescribes the  steepest rate of decay for the scalar  amplitudes  as
    \begin{equation}
    |\theta_m| \sim \mathcal{A}_m[\theta] \propto  \lambda^{-mh_{\max}}.
    \label{eq_Lmax2}
    \end{equation}
It follows that the sums with respect to $n$, which appear in Eq.~(\ref{eqS2}) and other similar expressions, converge exponentially in the inertial interval for any $\alpha < \lambda^{-2h_{\max}}$. 
In the numerics presented here, we have $\lambda^{-2h_{\max}} \approx  0.50$. This (finally!) substantiates the choices of $\alpha$ used in all our derivations and simulations. 

\corr
\section{Zero modes and hidden symmetry}
\label{secZMvcHS}
\rroc
\corr
The zero-mode and hidden symmetry approaches offer two complementary views on the origin of intermittency in scalar turbulence.
%The zero-mode and the hidden symmetry approaches provide two complementary visions on the intermittency at play in scalar turbulence. 
%including but not limited to Kraichnan-type dynamics.
%In this section, we will try to understand and compare the roles of zero-mode and hidden-symmetry approaches in understanding the passive scalar turbulence.
The  apparent advantage of the zero-mode theory is that it provides a quantitative approach, explicitly quantifying multifractality in terms of computed values of anomalous exponents.
%being the quantitative approach providing the values of anomalous exponents for structure functions. 
In  Kraichnan models, there are different ways to estimate zero modes, ranging from direct calculations based on fusion rules as  described in the present work~\cite{benzi1997analytic} to perturbation and renormalization-group techniques for the original continuous model~\cite{bernard1998slow,falkovich2001particles,kupiainen2007scaling}.  
In all these models the crucial simplification is
%which defines  Kraichnan dynamics is 
the replacement of the advecting Navier-Stokes velocity field  by a synthetic and shortly correlated-in-time (Gaussian) field. While this ensures that the correlation functions of the scalar field satisfy a closed, explicit  and solvable hierarchy of Hopf equations for equal-time correlations, it also latches the  computabitity of zero-mode solutions  on the white-in-time  nature of  Kraichnan-like dynamics.
%highlights the computabitity of zero-mode solutions  as a distinctive feature of Kraichnan-like dynamics.
%, which is not present in more realistic models of hydrodynamic turbulence. 
Still, footprints of zero modes as a mechanism underlying the phenomenon of intermittency were observed numerically both in passive scalar advection by a two-dimensional Navier-Stokes velocity field~\cite{celani2001statistical} and in hydrodynamic (nonlinear) shell models~\cite{arad2001statistical}. Unfortunately, a systematic zero-mode approach is missing in those cases and in otherwise more realistic models of turbulence. Increasing difficulties emerge when  addressing 
finite space-dimensionality \cite{chertkov1996anomalous}, non-Markov \cite{andersen1999shell,benzi2003intermittency,chaves2003lagrangian} or non-Gaussian advection \cite{peixoto2023spontaneous},  quenched statistics \cite{chaves2003lagrangian}, and nonlinearities \cite{thalabard2021inverse}.  
 Closed equations for single-time correlation functions are unlikely to exist for such realistic velocity fields, and  therefore generalizations of the  zero-mode approach would have to deal with multi-time correlation functions \cite{andersen1999shell}.

The hidden  symmetry approach explores a different, multiplicative nature of intermittency. It goes back to the Kolmogorov ideas of self-similarity~\cite{kolmogorov1941local,kolmogorov1962refinement} through the universality of so-called Kolmogorov multipliers~\cite{benzi1993intermittency,chen2003kolmogorov,mailybaev2022hiddenMT}. Unlike the original K41 theory, the approach uses a weaker but still exact hidden scaling symmetry of the ideal system. The rigorous formulation of the hidden symmetry hypothesis yields a first-principle approach, which does not rely on phenomenological assumptions but rather on  an explicit mathematical limit, here formulated as the limit $ \mathbb P^{(m)} \to \mathbb P^{\infty}$ for the hierarchy of stationnary measures associated to the rescaled system.
Consequences of the hidden symmetry for the KWB model proceed in an a  way very similar to previously studied  hydrodynamic models, in particular Sabra and Navier-Stokes dynamics~\cite{mailybaev2022shell,mailybaev2022hiddenMT}. 
This similarity is  seen already by comparing  the projective form of the hidden KWB dynamics \eqref{eq:rescaled} 
to the previously derived hidden Sabra and hidden Navier-Stokes systems --see also  Appendix \ref{app:B}.
More conceptually, it is rooted in commutation relations of the scaling symmetry group,  hence applying universally to all scale-invariant models --see \cite{mailybaev2022hidden}.
In hydrodynamic models, including the KWB model,  the  hidden scale invariance property currently stands as a conjecture, but our analysis provides  testable numerical evidence. The striking consequence is the derivation of the whole spectrum of multiscale properties (anomalous power laws, fusion rules, universality of Kolmogorov multipliers, multifractal spectrum, etc.) from the single property of hidden scale invariance. 

The derivations based on the hidden symmetry here  presented  are purely qualitative. For example, the hidden scale invariance yields the Perron-Frobenius scenario of the intermittency, where the anomalous exponents are obtained as eigenvalues. 
Only for very specific systems, those considerations are known to provide a quantitative theory  \cite{mailybaev2021solvable}, relying on the  solvability of the hidden dynamics. This is akin to zero-mode theory, which gives numerical values but only limited to Kraichnan models.
 We argue, however, that the two objects, the zero and Perron-Frobenius modes, are related. Indeed, the measures $\mu_p^{(m)}$ in the hidden-symmetric picture are generalizations (generators) of the correlation functions. As we showed in Section~\ref{sec:6}, all kinds of $p^{th}$-order correlation functions are obtained as averages of the measures $\mu_p^{(m)}$ for proper observables. 
For Kraichnan models, these averages satisfy the closed linear system of equations (eigenvalue problem), while the hidden scale invariance guarantees that such a system exists in a generalized form as the eigenvalue problem in terms of the measures. The latter is not limited to the Kraichnan model, and extends universally to a class of models sharing the same symmetry group. This observation suggests that the zero-mode and multiplicative mechanisms of intermittency are intrinsically connected, and a deeper understanding of this connection would be an interesting direction for future research.
\rroc

%###############################################
%CONCLUSION
\section{Concluding remarks}
\label{sec:conclusion}
By revisiting a simple model of  random scalar advection, we contrasted two  apparently distinct generic scenarios for spatial intermittency: zero-mode vs Perron-Frobenius associated to (hidden) statistical scale invariance.
The zero-mode scenario evidences  anomalous power-law scalings for  even-order structure functions through a direct, exact computation. 
By contrast, the Perron-Frobenius scenario relies on statistical symmetries obtained from the dynamical rescaling of the original scalar dynamics. While the rescaled dynamics is not linear any longer, what matters is that it possesses a weaker form of scale invariance: the hidden scale invariance, which we can evidence through statistical observations.
\corr
 The Perron-Frobenius scenario points towards a generic multiplicative origin for intermittency, which does not rely on the specific details of the dynamics but rather on its symmetries and nonlinearities.\rroc The fact that universality of multipliers is observed throughout fluid models, from the present KWB dynamics and shell models \cite{eyink2003gibbsian,vladimirova2021fibonacci,mailybaev2022shell} to direct simulations of fluid equations \cite{mailybaev2022hiddenMT} and experiments \cite{chen2003kolmogorov}, strongly suggests not only that the Perron-Frobenius scenario  is a generic mechanism behind fluid intermittency in fluid flows, but also that it can be evidenced in a systematic fashion.  %Let us here point out that the hidden KWB system \eqref{eq:rescaled} shares a structural analogy with the rescaled ideal systems obtained for non-linear shell models and the Navier-Stokes equations \cite{mailybaev2022hidden,mailybaev2022shell,mailybaev2022hiddenMT}, which can be traced back to structural analogies in the underlying dynamics.
Besides, while the hidden symmetry here discussed is limited to the inertial range of scales, extensions to include the  diffusion range are possible,  following the  ideas developed in \cite{mailybaev2023hidden} for turbulence models.
 
\corr Comparison of the zero-mode and the Perron-Frobenius scenario reveals intrinsic connections. The zero-mode approach reduces to an eigenvalue problem for correlation functions. Similarly, statistical recovery of hidden symmetry leads to an eigenvalue problem for specific measures, which can be viewed as generators of correlation functions. This potential connection between the zero-mode and the multiplicative nature of intermittency is a promising topic for future research.\rroc
%
%Compared to real flows, the  model considered in this paper combines several  simplifying features: lack of Galilean symmetry coming from the shell-model setting, and lack of intermittency for the carrier flow. Such features can be accomodated in the hidden symmetry framework using  slightly different rescaling strategies \cite{mailybaev2022hidden,mailybaev2022hiddenMT}. Natural extensions of this work
%include, but are not limited to, explicit address of real-space scalar  intermittency in various types of advecting environments, implications of hidden symmetry on multi-scale and multi-time correlators, and derivation of associated fusion rules. 

\subsection*{Acknowledgements}
\corr We thank Gregory Eyink for the discussions on the relation of hidden symmetry to fusion rules, and Chiara Calascibetta, Luca Biferale, Massimo Cencini and one anonymous reviewer for the discussions on the relation of hidden symmetry to zero-modes.\rroc
We acknowledge support  from the French-Brazilian network in Mathematics and from the Wolfgang Doeblin Research Federation
for  several monthly  visits  of S.T.  and A.A.M. respectively to IMPA and INPHYNI during the respective summer seasons of 2022-4. This work was also supported by the CNPq grant 308721/2021-7 and FAPERJ grant E-26/201.054/2022.

\subsection*{Data availability}
The numerical data featured  in this work  are available from the authors upon reasonable request.

\subsection*{Conflict of interest}
The authors declare they have no potential conflict of interest related to this work.
\newpage

%###############################################
%SUPPLEMENTAL
\appendix

%% APP 0
\section{\ito\, formulation of the shell model and numerical method}
\label{app:0}

\corr
Our technical derivations and numerics rely on  It\^o stochastic calculus. 
Using the Stratonovich to It\^o conversion formula \cite{oksendal2013stochastic} and boundary conditions (\ref{eq4}), one obtains the It\^o SDE version of Eq.~(\ref{eq:def_strato}) as  
%	\begin{equation}
 %	\label{eq5}
%	d \theta_n = \mathcal{N}_n[\theta,dw]
%	-\left(\frac{\gamma^{2n}+\gamma^{2n-2}(1-\delta_{n,1})}{2}+\mathrm{Pe}^{-1}\lambda^{2n}\right)\theta_n\,dt, 
%	\quad n \ge 1,
%	\end{equation}
	\begin{equation}
 	\label{eq5}
	d \theta_n =
	\mathcal{N}_n[\theta,dw]+
	(I_n+B_n-D_n) \theta_n\,dt, 
	\quad 1\le n \le n_{\max},
	\end{equation}
with  the shorthands  $\theta = \big(\theta_n\big)_{n \in \mathbb{Z}}$ and $dw = \big(dw_n\big)_{n \in \mathbb{Z}}$. %for bi-infinite sequences.
%where $\delta_{n,1}$ is the Kronecker delta and
The advection terms are 
	\begin{equation}
	\label{eq1_Bn}
	\mathcal{N}_n[\theta,dw] = \gamma^n \theta_{n+1}dw_n-\gamma^{n-1}\theta_{n-1}dw_{n-1}.
	\end{equation}
The drift terms  combine It\^o drifts, boundary  and diffusion terms, respectively:
\be
	\label{eq1_Ln}
	\begin{split}
%	& L_n=I_n+B_n-D_n,\\
%	&\text{for } 
	I_n
	= -\frac{\gamma^{2n} +\gamma^{2n-2}}{2}, \quad
	B_n = \dfrac{\delta_{n1}}{2}+\dfrac{\delta_{n n_{\max}}\gamma^{2n}}{2}, \quad
	D_n = \mathrm{Pe}^{-1}\lambda^{2n},
	\end{split}
\ee
where $\delta_{n1}$ and $\delta_{nn_{\max}}$ are the Kronecker deltas.
The coefficients $B_n$ retain the boundary effects at $n = 1$ and $n_{\max}$.
For the ideal model (\ref{eq5id}) the corresponding It\^o SDE version follows from Eq.~(\ref{eq5}) by neglecting the forcing and diffusion terms as
	\begin{equation}
 	\label{eq5ideal}
	d \theta_n =
	\mathcal{N}_n[\theta,dw]+
	I_n \theta_n\,dt, 
	\quad n \in \mathbb{Z}.
	\end{equation}
Our numerical simulations of the KWB dynamics rely on Milstein scheme \cite{higham2021introduction},
which 
%Despite the multidimensionality of noise, the scheme 
is here relatively simple to implement because the advection terms $\mathcal{N}_n[\theta,dw]$ are linear with respect to $\theta$. This scheme yields the increment at one time step $\Delta t$ as 
	\be
	\Delta \theta_n = \mathcal{N}_n(\theta, \Delta W) + (I_n+B_n-D_n)\theta_n \Delta t 
	+\,\dfrac{1}{2} \gamma^{2n} \theta_n \left( \Delta t- \Delta W_n^2 \right)+\gamma^{2n-2} \theta_n \left(\Delta t-\Delta W_{n-1}^2\right),
	\ee
where the $\Delta W_n$'s are  independent random numbers drawn from a centered Gaussian distribution $\sqrt{\Delta t} \;\mathcal N(0,1)$ with variance $\Delta t$. 
We use the integration step $\Delta t=5\times 10^{-7}$ and generate an ensemble of realizations in a steady state using a Monte-Carlo sampling as follows. Initially, we perform a run until time $T = 10^3$ and   then select the last time to seed $100$ different simulations,
which differ from each other by the realization of the noise. Each simulation is integrated until $T =10^3$
 with data output every time interval of $5\times 10^{-3}$, resulting in $2\times10^5$ outputs per realization.
Averages and densities are therefore  computed using $2 \times 10^8$ points per shell number. 
%Examples of times series and distributions are shown in Figure \ref{fig:1}(a).
\rroc
%% APP A
\section{\ito\, calculus for structure functions}
\label{app:A}

Here we derive equations for the second and fourth order structure functions.
The mutually independent Wiener processes $(w_n)_{n \in \mathbb{Z}}$ prescribe the fundamental property $dw_idw_i = \delta_{ij} dt$ \cite{pavliotis2014stochastic, 
oksendal2013stochastic, evans2012introduction}. Hence, using definition \eqref{eq1_Bn}, we derive
	\be
	\mathcal{N}_n[\theta,dw]\, \mathcal{N}_{n-j}[\theta,dw]= \mathcal{C}_{n(n-j)}[\theta]\, dt,\quad
	\mathcal{C}_{n(n-j)}[\theta] = 
	\begin{cases}
		  \gamma^{2n}   \left(  \theta_{n+1}^2 + \gamma^{-2}  \theta_{n-1}^2  \right),& j = 0; \\
		  -\gamma^{2n-2} \theta_n \theta_{n-1}, & j = 1; \\
		 0, & j > 1.
	\end{cases}
	\label{eq:L1}
	\ee
For the function $\theta_n^2$ with $n = 1,\ldots,n_{\max}$,  \ito's lemma~\cite{oksendal2013stochastic} with Eq.~(\ref{eq5}) yields
	\begin{equation}
	d \theta_n^2 = 2\theta_n \mathcal{N}_n[\theta,dw]+
	2 \left(I_n+B_n-D_n\right)\theta_n^2\,dt +\mathcal{C}_{nn}[\theta] \, dt,
	\label{eq_Ito_th2}
	\end{equation}
Let us average this equation over random realizations of the Wiener processes $w = (w_n)_{n \in \mathbb{Z}}$. The term $\av{\theta_n\mathcal{N}_n[\theta,dw]}$ vanishes, because $\theta$ and $dw$ are  mutually independent in \ito\, calculus. In the remaining terms, we substitute the coefficients from Eq.~(\ref{eq1_Ln}), the expression for $\mathcal{C}_{nn}[\theta]$ from Eq.~(\ref{eq:L1}), and use boundary conditions (\ref{eq4}). One can check that the resulting expression yields the scalar energy balance Eqs.~(\ref{eq7b}) and (\ref{eq8_Pi}).
	
Now, let us perform a similar analysis for any scalar function $\mathcal{F}[\theta]$. We assume that $\mathcal{F}[\theta]$ depends smoothly on shell scalars $\theta_n$ %from the inertial interval, which are
governed by the ideal model (\ref{eq5ideal}). Taking into account that $\mathcal{C}_{n(n-j)}[\theta] \equiv 0$ for $j > 1$ in Eq.~(\ref{eq:L1}), by \ito's lemma~\cite{oksendal2013stochastic} we have
	\begin{equation}
	d\mathcal{F}[\theta] = \sum_{j \in \mathbb Z } \left(
	\frac{\partial \mathcal{F}}{\partial \theta_j}\, d\theta_j
	+\frac{1}{2}\frac{\partial^2 \mathcal{F}}{\partial \theta_j^2}\,\mathcal{C}_{jj}[\theta]\, dt
	+\frac{\partial^2 \mathcal{F}}{\partial \theta_j\partial \theta_{j-1}}\,\mathcal{C}_{j(j-1)}[\theta]\, dt
	\right),\quad
	d\theta_j = \mathcal{N}_j[\theta,dw]+ I_j \theta_j\, dt.
	\label{eq_Ito_lemma}
	\end{equation}
Let us average Eq.~(\ref{eq_Ito_lemma}) over random realizations of $w$. Then the terms containing $\mathcal{N}_j[\theta,dw]$ vanish, and the remaining terms yield
\begin{comment}
 	\begin{equation}
	\frac{d}{dt}\av{\mathcal{F}[\theta]} = \sum_{j  \in \mathbb Z} \left\langle
	\frac{\partial \mathcal{F}}{\partial \theta_j}\, I_j \theta_j
	+\frac{1}{2}\frac{\partial^2 \mathcal{F}}{\partial \theta_j^2}\,\mathcal{C}_{jj}[\theta]
	+\frac{\partial^2 \mathcal{F}}{\partial \theta_j\partial \theta_{j-1}}\,\mathcal{C}_{j(j-1)}[\theta]
	\right\rangle.
	\label{eq_Ito_lemma_av}
	\end{equation}
\end{comment}
 	\begin{equation}
	\frac{d}{dt}\av{\mathcal{F}[\theta]} = \left\langle \mathcal L\left[ \mathcal F\right] \right\rangle,\quad \mathcal L[\mathcal F]:= \sum_{j  \in \mathbb Z} 	\frac{\partial \mathcal{F}}{\partial \theta_j}\, I_j \theta_j
	+\frac{1}{2}\frac{\partial^2 \mathcal{F}}{\partial \theta_j^2}\,\mathcal{C}_{jj}[\theta]
	+\frac{\partial^2 \mathcal{F}}{\partial \theta_j\partial \theta_{j-1}}\,\mathcal{C}_{j(j-1)}[\theta].
	\label{eq_Ito_lemma_av}
	\end{equation}

The functional $\mathcal L$ represents the (infinitesimal) generator of the scalar dynamics.
Equations for two-shell functions follow by taking $\mathcal{F}[\theta] = \theta^2_n\theta^2_{n+l}$ in  Eq.~\eqref{eq_Ito_lemma_av}. Elementary computations yield
	\be
	\dfrac{d}{dt}\av{ \theta_n^2\theta_{n+l}^2} = \av{2(I_n+I_{n+l}) \theta_n^2\theta_{n+l}^2}
	+ \left\{ \begin{array}{ll}
	6\av{\theta_{n}^2 \mathcal{C}_{nn}[\theta]}, & l = 0; \\
	4\av{\theta_{n}\theta_{n+1} \mathcal{C}_{(n+1)n}[\theta]}, & l = 1; \\
	\av{\theta_{n+l}^2 \mathcal{C}_{nn}[\theta] 
	+ \theta_{n}^2 \mathcal{C}_{(n+l)(n+l)}[\theta]}, & l \ge 2. \\
	\end{array}\right.
	\label{eq_4SFd}
	\ee
Substituting $I_n$ from Eq.~(\ref{eq1_Ln}), one obtains 

\corr
\be
\begin{split}
 \gamma^{-2n-l}\dfrac{d}{dt} \av{\theta^2_n\theta^2_{n+l}} 
=   -a_l \av{\theta_n^2\theta_{n+l}^2}  & + b_l \av{\theta_n^2\theta_{n+l+1}^2}+  b_l\gamma^{-2} \av{\theta_n^2\theta_{n+l-1}^2} \\
&  +  b_{-l} \av{\theta_{n+1}^2\theta_{n+l}^2}+ b_{-l} \gamma^{-2} \av{\theta_{n-1}^2\theta_{n+l}^2},
\end{split}
\label{eq:ds4}
\ee
with the coefficients \eqref{eqZ3}. System (\ref{eq:ds4B}) follows after averaging with respect to time.

\corr
\section{Computing the fourth-oder zero-mode solution}
\label{app_ZM4}

For large $l \gg 1$, we substitute coefficients (\ref{eqZ3}) and write Eq.~(\ref{eq:ds4BX}) asymptotically as
\be
 	\label{eqZM13}
	-(\gamma^l+\gamma^{l-2})C_l
	+\gamma^{l} C_{l+1}
	+\gamma^{l-2}C_{l-1}
	\approx 0,
	\quad l \gg 1,
\ee
where we neglected small terms with $\gamma^{-l}$. The asymptotic  linear system (\ref{eqZM13}) has two independent solutions defined (up to a constant factor) by $C_l \propto \gamma^{-2l}$ and $C_l \propto 1$. 
The  boundary condition \eqref{eqZM14} rules out the second solution.
The first solution yields a unique sequence $(C_l)_{l \ge 0}$ with $C_0=1$ solving the system \eqref{eq:ds4BX} for $l \ge 1$. Note that this solution depends on the yet unknown exponent $\zeta_4$. 
%
%We rule out the second solution, because it gives nondecreasing moments $S_{2,2}(k_n,k_{n+l})$ for large $l$. 
%Summarizing, we have the boundary conditions
%	\begin{equation} 	
%	\label{eqZM14}
%	C_0 = 1,\quad C_l \propto \gamma^{-2l},
%	\quad l \gg 1,
%	\end{equation}
%which select a unique sequence $C_l$, $l \ge 1$, solving system \eqref{eq:ds4BX} for a given exponent $\zeta_4$.
We now express $\lambda^{\zeta_4}$ from Eq.~\eqref{eq:ds4BX} for $l = 0$ with coefficients (\ref{eqZ3}) as
% one proceeds similarly but taking into account the equality $C_{-1} = C_1\lambda^{\zeta_4}$ following from Eqs.~(\ref{eq7B3}), (\ref{eqZ6}) and $S_{2,2}(k_n,k_{n-1}) = S_{2,2}(k_{n-1},k_n)$. This yields 
%	\be
%	-a_0 C_0  + 2b_0C_{1}(1+\gamma^{-2} \lambda^{\zeta_4}) = 0.
%	\label{eq:ds4BY}
%	\ee
%Substituting the coefficients from Eq.~(\ref{eqZ3}) and $C_0 = 1$ and solving with respect to $\lambda^{\zeta_4}$, one obtains
	\be
 	\label{eqZM12b}
	\lambda^{\zeta_4} = \frac{1+\gamma^{2}}{3 C_1}-\gamma^{2}.
	\ee
This is a nonlinear equation for the exponent $\zeta_4$ because $C_1$ depends on $\zeta_4$. 

Solution of Eq.~(\ref{eqZM12b}) can be found numerically with the following iteration procedure~\cite{benzi1997analytic}. One takes $\zeta_4 = 2\zeta_2$ as the initial guess and computes the corresponding solution $(C_l)_{l \ge 0}$ as described above. % and (\ref{eqZM14}). 
%{\color{red}This is done by ***SIMON:???***}\
% $C_1$ with backward iteration using Eq.~(\ref{eq:ds4BX}) and starting from 
% the asymptotic value (\ref{eqZM14}) taken for some $l \gg 1$. 
Then, the new corrected value of $\zeta_4$ is found from Eq.~(\ref{eqZM12b}). This process is iterated until the convergence is attained; see Fig. \ref{fig:2}(d).
In our numerics, taking $\xi=2/3$ we obtain $\zeta_4 \simeq 2.357$. 
%This exponent deviates from the monofractal prediction, i.e., $\zeta_4 \ne 2\zeta_2$. The deviation $\zeta_4-2\zeta_2 \simeq -0.309$ provides the scaling exponent for the flatness $S_2(\ell_n)/S_2^2(\ell_n) \propto \ell_n^{\zeta_4-2\zeta_2}$, which increases at small scales.
%As shown in Fig. \ref{fig:2}(c), the anomalous scaling for the flatness agrees with the zero-mode prediction over almost the full three decades of our numerics, with the  deviations seen at the ultraviolet end being caused by finite-size effects.
%Besides,  our numerics support the fusion rule ansatz  \eqref{eqZ6}. Fig. \ref{fig:2}(b) shows that the coefficients 
%$\sigma_{nl}/\sigma_{n0}$ become indeed independent of the index $n$ when both $n$ and $ n+l$ lie in inertial range, and feature the Kolmogorov scaling $C_l \propto \gamma^{-2l}$.
\rroc

\section{Derivation of the hidden KWB system}
\label{app:B}

In this section, we provide the detailed derivation of rescaled system \eqref{eq:rescaled}  from the ideal model Eq.~\eqref{eq5id}. For simplicity, we drop the superscript $(m)$ throughout this section.

Using expressions (\ref{eqS2}) and (\ref{eqS3}) for given $N$, we introduce the function $\mathcal{F}[\theta]$ as
	\begin{equation}
	\Theta_N = \mathcal{F}[\theta] := \frac{\theta_{m+N}}{\mathcal{A}_m[\theta]}, \quad
	\mathcal{A}_m[\theta] = \sqrt{\sum_{n \le m} \alpha^{m-n}\theta_n^2 }.
	\label{eq_F_Theta0}
	\end{equation}
Our intention is to use Eq.~(\ref{eq_Ito_lemma}). For the first derivatives, one computes
	\begin{equation}
	\frac{\partial\mathcal{F}}{\partial\theta_j}
	= \frac{\mathbbm{1}_{j = m+N}}{\mathcal{A}_m[\theta]}
	-\frac{\alpha^{m-j}\theta_{m+N} \theta_j}{\mathcal{A}_m^3[\theta]}
	\,\mathbbm{1}_{j \le m}.
	\label{eq_Ito_F1}
	\end{equation}
Considering $N > 1$, for the second derivatives we have
	\begin{equation}
	\frac{\partial^2\mathcal{F}}{\partial\theta_j^2} 
	= 
	-\frac{2\alpha^{m-j} \theta_j}{\mathcal{A}_m^3[\theta]}
	\,\mathbbm{1}_{j = m+N \le m}
	+\frac{\alpha^{m-j}\theta_{m+N}}{\mathcal{A}_m^3[\theta]} \left(
	\frac{3\alpha^{m-j}\theta_{j}^2}{\mathcal{A}_m^2[\theta]}-1 \right) 
	\mathbbm{1}_{j \le m},
	\label{eq_Ito_F2a}
	\end{equation}
	\begin{equation}
	\frac{\partial^2\mathcal{F}}{\partial\theta_j\partial\theta_{j-1}} 
	= 
	-\frac{\alpha^{m-j+1} \theta_{j-1}}{\mathcal{A}_m^3[\theta]}
	\,\mathbbm{1}_{j = m+N \le m+1}
	-\frac{\alpha^{m-j} \theta_{j}}{\mathcal{A}_m^3[\theta]}
	\,\mathbbm{1}_{j = m+N+1 \le m}
	+\frac{3\alpha^{2m-2j+1}\theta_{m+N}\theta_{j}\theta_{j-1}}{\mathcal{A}_m^5[\theta]} 
	\,\mathbbm{1}_{j \le m}.
	\label{eq_Ito_F2b}
	\end{equation}
Substituting Eqs.~(\ref{eq_Ito_F1})--(\ref{eq_Ito_F2b}) into (\ref{eq_Ito_lemma}), after elementary manipulations we have
	\begin{equation}
	\begin{split}
	& d\Theta_N = d\mathcal{F}[\theta] = 
	\frac{d\theta_{m+N}}{\mathcal{A}_m[\theta]}
	-\sum_{j \le m} \alpha^{m-j}
	\frac{\theta_{m+N}\theta_j}{\mathcal{A}_m^2[\theta]} 
	\frac{d\theta_j}{\mathcal{A}_m[\theta]} 
	\\[7pt]
	& 
	-\alpha^{-N}\frac{\theta_{m+N}}{\mathcal{A}_m[\theta]}
	\frac{\mathcal{C}_{(m+N)(m+N)}[\theta]}{\mathcal{A}_m^2[\theta]} \, dt
	\,\mathbbm{1}_{N \le 0}
	+\frac{1}{2}\sum_{j \le m} \alpha^{m-j}
	\frac{\theta_{m+N}}{\mathcal{A}_m[\theta]} \left(
	3\alpha^{m-j} \frac{\theta_{j}^2}{\mathcal{A}_m^2[\theta]}-1 \right)
	\frac{\mathcal{C}_{jj}[\theta]}{\mathcal{A}_m^2[\theta]} \, dt
	\\[7pt]
	& 
	-\alpha^{-N+1}\frac{\theta_{m+N-1}}{\mathcal{A}_m[\theta]}
	\frac{\mathcal{C}_{(m+N)(m+N-1)}[\theta]}{\mathcal{A}_m^2[\theta]} \, dt
	\,\mathbbm{1}_{N \le 1}
	-\alpha^{-N-1}\frac{\theta_{m+N+1}}{\mathcal{A}_m[\theta]}
	\frac{\mathcal{C}_{(m+N+1)(m+N)}[\theta]}{\mathcal{A}_m^2[\theta]} \, dt
	\,\mathbbm{1}_{N \le -1}
	\\[7pt]
	& 
	+\sum_{j \le m} 3\alpha^{2m-2j+1}
	\frac{\theta_{m+N}\theta_{j}\theta_{j-1}}{\mathcal{A}_m^3[\theta]}
	\frac{\mathcal{C}_{j(j-1)}[\theta]}{\mathcal{A}_m^2[\theta]} \, dt.
	\end{split}
	\label{eq_Ito_Theta}
	\end{equation}
Using expressions (\ref{eqS3}) and (\ref{eq:L1}), one writes
	\begin{equation}
	\frac{\mathcal{C}_{jj}[\theta]}{\mathcal{A}_m^2[\theta]} \, dt = 
	\gamma^{2J}   \left(  \Theta_{J+1}^2 + \gamma^{-2}  \Theta_{J-1}^2  \right)
	d\tau,
	\quad
	\frac{\mathcal{C}_{j(j-1)}[\theta]}{\mathcal{A}_m^2[\theta]} \, dt = 
	-\gamma^{2J-2}  \Theta_J \Theta_{J-1}
	d\tau,
	\quad
	j = m+J.
	\label{eq_Ito_Theta2}
	\end{equation}
Similarly, using expressions (\ref{eqS3}), (\ref{eq5ideal}), (\ref{eq1_Bn}) and (\ref{eq1_Ln}), we derive 
	\begin{equation}
	\frac{d\theta_j}{\mathcal{A}_m[\theta]}
	= \mathcal{N}_J[\Theta,dW]+I_J \Theta_J d\tau,
	\quad
	j = m+J.
	\label{eq_Ito_Theta2b}
	\end{equation}
Substituting Eqs.~(\ref{eq_Ito_Theta2}) and (\ref{eq_Ito_Theta2b}) into (\ref{eq_Ito_Theta}), using expressions (\ref{eqS3}) and changing the summation index as $J = j-m$ yields	
    \begin{equation}
	\begin{split}
	d\Theta_N = & \,\mathcal{N}_N[\Theta,dW]+I_N\Theta_N d\tau
	-\Theta_{N}\sum_{J \le 0} \alpha^{-J} \Theta_J 
	\left(\mathcal{N}_J[\Theta,dW]+I_J\Theta_J d\tau\right)
	 \\[3pt]
	 &
 	-\alpha^{-N}\Theta_{N}
	\gamma^{2N}   \left(  \Theta_{N+1}^2 + \gamma^{-2}  \Theta_{N-1}^2  \right)
	d\tau
	\,\mathbbm{1}_{N \le 0}
	 \\[3pt]
	 &
	+\alpha^{-N+1}
	\gamma^{2N-2}  \Theta_N \Theta_{N-1}^2
	d\tau
	\,\mathbbm{1}_{N \le 1}
	+\alpha^{-N-1}
	\gamma^{2N}  \Theta_{N+1}^2 \Theta_{N}
	d\tau
	\,\mathbbm{1}_{N \le -1}
	 \\[3pt]
	& +\Theta_{N}\sum_{J \le 0} 
	\left[\frac{\alpha^{-J}}{2}
	(3\alpha^{-J}\Theta_J^2-1) \gamma^{2J} 
	 (  \Theta_{J+1}^2 + \gamma^{-2}  \Theta_{J-1}^2  ) 
	 -3 \alpha^{1-2J}\gamma^{2J-2}\Theta_J^2\Theta_{J-1}^2 
	 \right] d\tau. 
	\end{split}
	\label{eq_Ito_Theta3}
	\end{equation}
Equation~\eqref{eq_Ito_Theta3}  is the \ito\, version of the hidden KWB dynamics.

\corr
Let us show that Eq.~\eqref{eq:rescaled} has the more compact form
\be
\label{eq:rescaled_A1}
 d\Theta_N^{(m)} = \circ \Lambda_N\left[\Theta^{(m)},
 \mN\big[\Theta^{(m)},dW^{(m)}\big] \right],
 \quad N\in \mathbb Z,
\ee
where we introduced the nonlinear operator
\be
\label{eq:rescaled_Lambda}
\Lambda_N[\Theta,V]= V_N-\Theta_N \sum_{J\le0} \alpha^{-J}\Theta_J V_J.
\ee
Here $\circ$ denotes mid-point evaluation of the right-hand-side through Stratonovich Rule.
\rroc
Then, Eq.~(\ref{eq_Ito_Theta3}) is recovered from  Eqs.~(\ref{eq:rescaled_A1}) and (\ref{eq:rescaled_Lambda})
 by expanding to order $d\tau$ the following practical Stratonovich-\ito\,  conversion formula \cite{evans2012introduction}
\begin{equation}
%\label{eq:hiddenito-proj0}
   \circ  \Lambda_N\left[\Theta,\mN\left[\Theta,dW\right]\right] = \Lambda_N\left[\Theta+\frac{d\Theta} 2,\mN\left[\Theta+\frac{d\Theta}2,dW\right]\right],
\end{equation}
yielding 
\begin{equation}
\circ\Lambda_N[\Theta,\mathcal N] = \Lambda_N[\Theta,\mN+I \Theta d\tau]-\mN_N d\mA\left[\Theta,\mN\right]+\dfrac{\Theta_N}{2} \Big( 3 \left(d\mA\left[\Theta,\mN\right]\right)^2   -d\mA\left[\mN,\mN\right]\Big)
\label{eq:hiddenito}
\end{equation}
with the  shorthands
\begin{equation}
    \mN:=\mathcal N[\Theta,dW],\quad d\mA[V , W]:=\sum_{J\le 0} \alpha^{-J}V_JW_J.
\end{equation}
The first right-hand side term in Eq.~\eqref{eq:hiddenito} provides the first line in Eq.~\eqref{eq_Ito_Theta3}. The remaining terms are generalized \ito\, terms, which upon using Eq.~\eqref{eq:L1},  precisely yield the remaining lines in  Eq.~\eqref{eq_Ito_Theta3}.
\corr Using the explicit form of nonlinear terms (\ref{eq1_Bn}) in Eqs.~(\ref{eq:rescaled_A1}) and (\ref{eq:rescaled_Lambda}) completes the derivation of the Hidden KWB dynamics \eqref{eq:rescaled}.\rroc

\section{Derivation of hidden symmetry transformation}
\label{app:C}

Here we prove that relations (\ref{eq:HS}) and (\ref{eq:apm}) map the processes $\Theta^{(m)}(\tau^{(m)})$ and $W^{(m)}(\tau^{(m)})$ into $\Theta^{(m\pm1)}(\tau^{(m\pm1)})$ and $W^{(m\pm1)}(\tau^{(m\pm1)})$. These processes satisfy equations of the same rescaled system (\ref{eq:rescaled}), which is independent of $m$. Therefore, it follows that the transformations (\ref{eq:HS}) induce a group of symmetries of the rescaled equations. 

One can deduce relations (\ref{eq:HS}) and (\ref{eq:apm}) for $W$ and $\tau$ directly from Eq.~(\ref{eqS3}) considered for $m$ and $m \pm 1$. It remains to show that the operators $a^{\pm 1}$ map $\Theta^{(m)}$ into $\Theta^{(m\pm1)}$. Let us first consider the case of positive sign. Using (\ref{eqS2}), we write
	\begin{equation}
 	\label{eqA3_1}
	\mathcal{A}_{m+1}[\theta] = \sqrt{\theta_{m+1}^2+\alpha \theta_{m}^2+\alpha^2 \theta_{m-1}^2+\cdots}
	= \sqrt{\theta_{m+1}^2+\alpha \mathcal{A}_{m}^2[\theta] }
	= \mathcal{A}_{m}[\theta] \sqrt{\left(\Theta_{1}^{(m)}\right)^2+\alpha },
	\end{equation}
where we expressed $\Theta_{1}^{(m)} = \theta_{m+1}/\mathcal{A}_m[\theta]$ in the last equality from Eq.~(\ref{eqS3}).
Using Eq.~(\ref{eqA3_1}) in (\ref{eqS3}), we write
	\begin{equation}
 	\label{eqA3_2}
	\Theta_N^{(m+1)} = \frac{\theta_{m+N+1}}{\mathcal{A}_{m+1}[\theta]}
	= \frac{\theta_{m+N+1}}{\mathcal{A}_{m}[\theta] \sqrt{\Theta_{1}^2+\alpha }}
	= \frac{\Theta_{N+1}^{(m)}}{\sqrt{\alpha+\left(\Theta_{1}^{(m)}\right)^2 }},
	\end{equation}
which yields the first relation in Eq.~(\ref{eq:apm}). Similarly, for the negative sign, we have 
	\begin{equation}
 	\label{eqA3_3}
	\mathcal{A}_{m-1}[\theta] = \sqrt{\theta_{m-1}^2+\alpha \theta_{m-2}^2+\cdots}
	= \sqrt{(\mathcal{A}_{m}^2[\theta]-\theta_{m}^2)/\alpha}
	= \mathcal{A}_{m}[\theta] \sqrt{\frac{ 1-\left(\Theta_{0}^{(m)}\right)^2 }{\alpha}}.
	\end{equation}
Then we derive the second relation in Eq.~(\ref{eq:apm}) as
	\begin{equation}
 	\label{eqA3_4}
	\Theta_N^{(m-1)} = \frac{\theta_{m+N-1}}{\mathcal{A}_{m-1}[\theta]}
	= \frac{\theta_{m+N-1}}{\mathcal{A}_{m}[\theta]}
	\sqrt{\frac{\alpha}{ 1-\left(\Theta_{0}^{(m)}\right)^2 }}
	= \sqrt{\frac{\alpha}{ 1-\left(\Theta_{0}^{(m)}\right)^2 }}\,\Theta_{N-1}^{(m)}.
	\end{equation}

%-------------------------------------------------------------- 
%-------------------------------------------------------------- 
\section{Zero modes as statistical conservation laws}
\label{app:D}
We here point out that  the zero-modes solving the eigenvalue problems 
\eqref{eq7B1_P} and \eqref{eqZM_EP4}
%\eqref{eq:M2} and	\eqref{eq:M4}
 relate to the presence of  explicit statistical conservations laws, which we respectively label as 
$\Gamma_2$ and $\Gamma_4$.  

%-------------------------------------------------------------- 
\subsection{ The $\Gamma_2$ invariant}
Let us show that the Kolmogorov zero-mode $\mathcal S_2(\ell_n) = \gamma^{-2n}$ solving Eq.~\eqref{eq7B1} is dual to the ideal conservation of the statistical invariant
\be
	\Gamma_2 := \sum_{n \in \mathbb Z}\gamma^{-2n} \av{\theta_n^2}.
 %=  \mS_2^T \av{\theta^2},
	\label{eq:G2}
\ee
%with $\mS_2^T$ denoting the transposed (row-vector) sequence.
The ideal conservation of $\Gamma_2$  associated to Eq.~\eqref{eq7b_Id}
 is seen from the direct telescopic calculation
\be
	\dfrac{d}{dt} \Gamma_2  = \sum_{n \in \mathbb Z}\left(\gamma^{-2}\av{\theta_{n-1}^2}   -(\gamma^{-2}+1) \av{\theta_n^2}+ \av{\theta_{n+1}^2} \right) =0,
%\sum_n \gamma^{-2n} (-\gamma^{2n}(1+\gamma^{-2}) \av{\theta_n^2}+\gamma^{2n}\av{\theta_{n+1}^2}+\gamma^{2n-2}\av{\theta_{n+1}^2})
\ee
where the last equality follows after the formal change of indices $n \pm 1 \mapsto n$.
More fundamentally, let us write Eq.~\eqref{eq7b_Id} as
\be
	\dfrac{d\left\langle \theta_n^2\right\rangle}{dt}  = \sum_{n' \in \mathbb Z} M_{n n'}  \left\langle \theta_{n'}^{2}\right\rangle 
\ee
with the matrix elements
\be
 M_{nn'} = 
 \gamma^{2n-2}\delta_{(n-1)n'}
 -\left(\gamma^{2n-2}+\gamma^{2n} \right)\delta_{nn'}+\gamma^{2n}\delta_{(n+1)n'}.
\ee
The stationary solution $\left\langle \theta_n^2\right\rangle = \gamma^{-2n}$ is a right zero-mode of $M$, i.e. $\sum_{n' \in \mathbb Z}
M_{nn'} \gamma^{-2n'} =0$ for all $n \in \mathbb Z$.
Notice that the matrix $M = (M_{n n'})$ is symmetric. Hence, the ideal conservation of $\Gamma_2$ given by Eq.~(\ref{eq:G2}) follows from $\gamma^{-2n}$ being also a left zero-mode, i.e.
$\sum_{n \in \mathbb Z}
\gamma^{-2n}
M_{nn'} =0$ for all $n' \in \mathbb Z$.
In that sense, the ideal statistical conservation of  $\Gamma_2$ is dual to the Kolmogorov scaling $\mathcal S_2(\ell_n) = \gamma^{-2n}$.

%-------------------------------------------------------------- 
\subsection{The $\Gamma_4$ invariant.}
The same  type of duality holds for the fourth-order zero-mode: $\sigma_{nl} = C_l S_4(\ell_n)$ with $S_4(\ell_n) \propto \ell^{\zeta_4}$ from Section~\ref{subsec_S4}. This zero-mode ties to the statistical conservation  of the quantity
\be
	\label{eq:G4}
	\Gamma_4 = \sum_{n} g_n,\quad g_n:=\sum_{l \ge 0} c_l r^n \av{\theta_n^2 \theta_{n+l}^2},
\ee
where $c_l=6C_l-5\delta_{l0}$ and $r=\lambda^{-\zeta_4}$. This conservation law can be expressed in the form
\be
	\dfrac{d}{dt} g_n = \Pi^g_{n-1}-\Pi^g_n,
	\label{eq:G4balance1}
\ee
where some straightforward but tedious algebra (see the next \S \ref{app:A3}) yields the flux as
\be
	\Pi^{g}_n = \sum_{l\ge0} \pi^g_{ln},\quad
	\pi^g_{ln}:=
	\begin{cases}
	-\left(r \gamma^{2}\right)^{n}   \av{\theta_{n+1}^4}, & 
 l= 0; \\
	\left(r \gamma^{2}\right)^{n}\left(   c_{2}  \av{\theta_{n+1}^2\theta_{n+2}^2}- 6r\av{\theta_{n}^2\theta_{n+1}^2} \right), & 
 l= 1; \\
	\left( r\gamma^{2}\right)^{n} \left(c_{l+1} \av{\theta_{n+1}^2\theta_{n+l+1}^2}- c_{l-1} r\av{\theta_{n}^2\theta_{n+l}^2}\right), 
    & 
    l\ge 1.
	\end{cases}
	\label{eq:G4balance2}
\ee

The dual origin of the invariant $\Gamma_4$ can be seen by writing Eq.~(\ref{eq:ds4}) as
\be
\begin{split}
 \gamma^{-2n-l}\dfrac{d}{dt} \av{\theta^2_n\theta^2_{n+l}} 
=   
\sum_{n',l'} {\mathcal M}_{nl,n'l'} \av{\theta^2_{n'}\theta^2_{n'+l'}}.
\end{split}
\label{eq:ds4App}
\ee
Here the right-hand side vanishes for the fourth-order zero-mode solution $\left\langle \theta_n^2\theta_{n+l}^2\right\rangle = \sigma_{nl}$. Let $\tilde{\sigma}_{nl}$ be the corresponding left zero-mode satisfying the equation $\sum_{n,l} \tilde{\sigma}_{nl}{\mathcal M}_{nl,n'l'} = 0$ for all $n'$ and $l'$. One can show that the left zero-mode is found explicitly as $\gamma^{-2n-l}\tilde{\sigma}_{nl} = \left(6-5\delta_{l0}\right) \sigma_{nl} = c_lr^n$. This dual zero-mode yields the conservation law for $\Gamma_4$ in Eq.~(\ref{eq:G4}).

%-------------------------------------------------------------- 
\subsection{Technical derivation of the  $\Gamma_4$ flux in Eq. \eqref{eq:G4balance2}}  \label{app:A3}
We first observe
%
%we first recall the definition of the coefficient $C_l$'s as :
%	\be
 %	\label{eqZM12_aux}
%	\left\lbrace 
%	\begin{split}
%		-a_0 + 6 C_1 \left(1+ \gamma^{-2} \lambda^{\zeta_4}\right)  = 0
%	 \hspace{1cm} & (l=0)\\
%		      C_{2}\left( \gamma^2 + \gamma^{-2} \lambda^{\zeta_4} \right) - \gamma a_1 C_1  +  1+  \lambda^{-\zeta_4} =0	 \hspace{1cm} & (l=1)\\
%	      C_{l+1}\left( \gamma^l + \gamma^{-l-2} \lambda^{\zeta_4} \right) -a_l C_l  +  C_{l-1} \left(\gamma^{l-2}     + \gamma^{-l}  \lambda^{-\zeta_4} \right) =0	 \hspace{1cm} & (l\ge 2)
%	\end{split}\right.
%	\ee
%
that the  coefficients  $c_l=6C_l-5\delta_{l,0}$ and $r=\lambda^{-\zeta_4}$ satisfy
% $r$ and $c_l,l\ge 0 $   as
	\be
 	\label{eqZM13_aux}
	\left\lbrace 
	\begin{split}
		-a_0 + c_1 \left(1+ \gamma^{-2} /r\right)  = 0
	 \hspace{1cm} & (l=0)\\
	      c_{2}\left( \gamma^2 + \gamma^{-2}/r \right) -a_1\gamma c_1 +   6+  6r  =0	 \hspace{1cm} & (l= 1)\\
	      c_{l+1}\left( \gamma^l + \gamma^{-l-2}/r \right) -a_l c_l  +  c_{l-1} \left(\gamma^{l-2}     + \gamma^{-l}  r \right) =0	 \hspace{1cm} & (l\ge 2)
	\end{split}\right.
	\ee
with the coefficients of Eq.~\eqref{eqZ3}.
\begin{comment}
\be
a_l=\begin{cases}
&2+2\gamma^{-2} \hspace{0.2cm}(l=0)\\
&\gamma+\gamma^{-3}+6\gamma^{-1} \hspace{0.2cm}  (l=1)\;\\
&\gamma^l+\gamma^{-l-2}+\gamma^{-l}+ \gamma^{l-2}  \hspace{0.2cm} (l \ge 2)
\end{cases} \text{and }
b_l=\begin{cases}
&3 \hspace{0.2cm}(l=0)\\
&\gamma^l  \;(otherwise)
\end{cases}
\ee
\end{comment}
%
From Eq. \eqref{eq:ds4} and  the definition \eqref{eq:G4}, we now write
\be
	\label{eq:G4a}
	\begin{split}
	 \dfrac{d}{dt}g_n  
%&= \sum_{l \ge 0} c_l r^n \dfrac{d}{dt}\av{\theta_n^2 \theta_{n+l}^2}\\	 
  & =  \sum_{l \ge 0} c_l r^n\gamma^{2n+l} \left(  -a_l \av{\theta_n^2\theta_{n+l}^2}   + b_l \av{\theta_n^2\theta_{n+l+1}^2}+ b_l \gamma^{-2} \av{\theta_n^2\theta_{n+l-1}^2}\right) \\
&\quad \quad \quad \quad   \;\;+  c_lr^n\gamma^{2n+l}\left(b_{-l}  \av{\theta_{n+1}^2\theta_{n+l}^2}+ b_{-l} \gamma^{-2} \av{\theta_{n-1}^2\theta_{n+l}^2}\right)\\
%%%%%%%%%
%  = &\sum_{l\ge 0}  -a_l  \gamma^{2n+l}c_lr^n   \av{\theta_n^2\theta_{n+l}^2}   + b_l \gamma^{2n+l}c_lr^n  \av{\theta_n^2\theta_{n+l+1}^2}+ b_l \gamma^{-2} \gamma^{2n+l}c_lr^n \av{\theta_n^2\theta_{n+l-1}^2} \\
%&  +  \gamma^{2n+l} c_lr^nb_{-l}  \av{\theta_{n+1}^2\theta_{n+l}^2}+  \gamma^{2n+l} c_lr^nb_{-l}  \gamma^{-2} \av{\theta_{n-1}^2\theta_{n+l}^2}\\
%%%%%%%%%
%  & = \sum_{l\ge 0}  -a_l  \gamma^{2n+l}c_lr^n   \av{\theta_n^2\theta_{n+l}^2}   + \sum_{l\ge 1} b_{l-1} \gamma^{2n+l-1}c_{l-1}r^n  \av{\theta_n^2\theta_{n+l}^2}+ \sum_{l\ge -1} b_{l+1} \gamma^{-2} \gamma^{2n+l+1}c_{l+1}r^n \av{\theta_n^2\theta_{n+l}^2} \\
%&  \;\; +\sum_{l\ge -1}   \gamma^{2n+l+1} c_{l+1}r^nb_{-l-1}  \av{\theta_{n+1}^2\theta_{n+l+1}^2}+ \sum_{l\ge 1}  \gamma^{2n+l-1} c_{l-1}r^nb_{-l+1}  \gamma^{-2} \av{\theta_{n-1}^2\theta_{n+l-1}^2}\\
%%%%%%%%%
% & = \sum_{l\ge 2}   \gamma^{2n+l}r^n  \av{\theta_n^2\theta_{n+l}^2} \left( -a_l c_l    +  b_{l-1} \gamma^{-1}c_{l-1}   + b_{l+1} \gamma^{-1}  c_{l+1}\right) \\
%& \;\; +\sum_{l\ge 2}   \gamma^{2n+l+1} c_{l+1}r^nb_{-l-1}  \av{\theta_{n+1}^2\theta_{n+l+1}^2}+  \gamma^{2n+l-1} c_{l-1}r^nb_{-l+1}  \gamma^{-2} \av{\theta_{n-1}^2\theta_{n+l-1}^2} 
 %+BT\\
%%%%%%%%%
  & = BT+\sum_{l\ge 2}   \gamma^{2n+l}r^n  \av{\theta_n^2\theta_{n+l}^2} \overbrace{\left( -a_l c_l    +   \gamma^{l-2}c_{l-1}   + \gamma^{l}  c_{l+1}\right)}^{-\gamma^{-l-2}r^{-1}c_{l+1} -\gamma^{-l}rc_{l-1}} \\
&\quad \quad  \quad+\sum_{l\ge 2}   \gamma^{2n} c_{l+1}r^n  \av{\theta_{n+1}^2\theta_{n+l+1}^2}+  \gamma^{2n} c_{l-1}r^n \gamma^{-2} \av{\theta_{n-1}^2\theta_{n+l-1}^2} \\
%&\;\;\; +BT\\
&=  BT +\sum_{l \ge 2} \pi^g_{l(n-1)}-\pi^g_{ln},\\%\quad  \text{ with } \pi_{l,n}:=\left( r\gamma^{2}\right)^{n} \left(c_{l+1} \av{\theta_{n+1}^2\theta_{n+l+1}^2}- c_{l-1} r\av{\theta_{n}^2\theta_{n+l}^2}\right)\\
%\end{split}	 
%\ee
%with 
%\be
%\begin{split}
\text{with }BT%& =  %\sum_{0,1}  -a_l  \gamma^{2n+l}c_lr^n   \av{\theta_n^2\theta_{n+l}^2}   + \sum_{1} b_{l-1} \gamma^{2n+l-1}c_{l-1}r^n  \av{\theta_n^2\theta_{n+l}^2}+ \sum_{-1,0,1} b_{l+1} \gamma^{-2} \gamma^{2n+l+1}c_{l+1}r^n \av{\theta_n^2\theta_{n+l}^2} \\
%&  +\sum_{ -1,0,1}   \gamma^{2n+l+1} c_{l+1}r^nb_{-l-1}  \av{\theta_{n+1}^2\theta_{n+l+1}^2}+ \sum_{1}  \gamma^{2n+l-1} c_{l-1}r^nb_{-l+1}  \gamma^{-2} \av{\theta_{n-1}^2\theta_{n+l-1}^2}\\
%%%%%%%%%%%%%%%%%%%%%%%%%%%%%%%%%%
& =    \gamma^{2n} r^n   \av{\theta_n^4} \left( -a_0 + c_1 \right)+    \gamma^{2n} r^n  c_{1} \av{\theta_{n+1}^4}\\
%%%%%%%%%%%%%%%%%%%%%%%%%%%%%%%%%%%%%%%%%%
& \;\; +  \gamma^{2n}r^n   \av{\theta_n^2\theta_{n+1}^2}\underbrace{\left( -a_1c_1 \gamma  +  6  + \gamma ^2 c_{2} \right)}_{-6r-6\gamma^{-2}r^{-1}c_2}
 +  6  \gamma^{2n-2} r^n \av{\theta_n^2\theta_{n-1}^2}   
  +   \gamma^{2n} r^n c_{2}  \av{\theta_{n+1}^2\theta_{n+2}^2}\\
%%%%%%%%%%%%%%%%%%%%%%%%%%%%%%%%%%
& =   \sum_{l=0,1} \pi^g_{l(n-1)}-\pi^g_{ln}.  \\
%& \text{with }\pi_0(n):=-\gamma^{2n-2} r^{n-1}   \av{\theta_n^4}  \;\;  \pi_1(n):=-  6  \gamma^{2n-2} r^n \av{\theta_n^2\theta_{n-1}^2} +  \gamma^{2n-2} r^{n-1} c_{2}  \av{\theta_{n+1}^2\theta_{n+2}^2}\\
%& \text{for }
%\quad \pi_{0,n}:=-\left(r \gamma^{2}\right)^{n}   \av{\theta_{n+1}^4}, \quad  \pi_{1,n}:=  \left(r \gamma^{2}\right)^{n}\left(   c_{2}  \av{\theta_{n+1}^2\theta_{n+2}^2}- 6r\av{\theta_{n+1}^2\theta_{n}^2} \right);
\end{split}
\ee
This completes the derivation of Eqs.~\eqref{eq:G4balance1}--\eqref{eq:G4balance2}.

\newcommand{\il}{l}
\newcommand{\ip}{1 \le p \le m}
\newcommand{\ilp}{1 \le l,p \le m}

%###################################################
%BIBLIO
\bibliography{biblio}
%###################################################
\end{document}